%% file: MultiJets2011.tex
\RequirePackage{lineno}
\documentclass[final,5p,times,twocolumn]{notelsarticle}
\pdfoutput=1 
\usepackage{atlasphysics}
\usepackage[bf,tight]{subfigure}
\usepackage{mathrsfs}
\usepackage{graphicx}
\usepackage{multirow}
\usepackage{rotating}
\usepackage{amsmath}
\usepackage{url}
\usepackage{textpos}
\usepackage{hyperref}

\biboptions{sort,merge}

\input{susydefs}
\graphicspath{./}
\DeclareGraphicsExtensions{.pdf}

\newlength{\plotwidth}
\newcommand{\maybesection}[1]{\section{#1}}
\newcommand{\ourintlumi}{{1.34\,\ifb{}}}
\newcommand{\mSUGRA}{MSUGRA/CMSSM}

\newcommand{\ourpt}{p_\mathrm{T}}
\newcommand{\mT}{\ensuremath{m_{\rm T}}}
\newcommand{\anactualptmissvecwithapinit}{{\vec{P}}_\mathrm{T}^\mathrm{miss}}
\newcommand{\ourvecptmiss}{\ensuremath{\anactualptmissvecwithapinit}}
\newcommand{\ourmagptmiss}{{\ensuremath \MET}}
\renewcommand{\HT}{{\ensuremath{H_{\rm T}}}}
\newcommand{\njet}{{\ensuremath{N_{\rm jet}}}}
\newcommand{\metsig}{{\ensuremath{\ourmagptmiss/\sqrt{\HT}}}}
\newcommand{\rootgev}{{\ensuremath{\GeV^{1/2}}}}

\newcommand{\sigmamaxbsm}{\ensuremath{\sigma_{\rm BSM,max}^{95\%}}}
\newcommand{\nmaxbsm}{\ensuremath{N_{\rm BSM,max}^{95\%}}}
\newcommand{\alpgen}{{\tt ALPGEN}}
\newcommand{\herwig}{{\tt HERWIG}}

\newcommand{\jimmy}{{\tt JIMMY}}
\newcommand{\prospino}{{\tt PROSPINO}}
\newcommand{\cteqVER}[1]{{\tt CTEQ#1}}

\newcommand{\msugra}{MSUGRA/CMSSM}
\newcommand{\leptonic}{`leptonic'} 
\newcommand{\citemsugraandcmssm}{\cite{Chamseddine:1982jx,*Barbieri:1982eh,*Ibanez:1982ee,*Hall:1983iz,*Ohta:1982wn,*Kane:1993td}}
\newcommand{\topcr}{{\ensuremath{{CR}^{\ttbar}_{j,p}}}}

\newcommand{\oursubref}[1]{{\bf (\subref{#1})}}
\newcommand{\ourtabref}[1]{Table~\ref{#1}}
\newcommand{\ourfigref}[1]{Figure~\ref{#1}}
\newcommand{\oursecref}[1]{Section~\ref{#1}}

\newcommand{\multijettitle}{Search for new phenomena in final states with 
large jet multiplicities and missing transverse momentum 
using $\sqrt{s}=7\TeV$ pp collisions with the ATLAS detector}

\begin{document}

\title{\multijettitle}
\author{The ATLAS Collaboration}

\begin{abstract}
\begin{textblock}{6}[0,1](0,-3){\small CERN-PH-EP-2011-155, Submitted to JHEP}
\end{textblock}
Results are presented of a search for any particle(s) decaying to 
six or more jets in association with missing transverse momentum.
The search is performed using \ourintlumi{} of $\sqrt{s}=7$\,\TeV{} 
proton-proton collisions recorded by the ATLAS detector during 2011. 
Data-driven techniques are used to determine the backgrounds in 
kinematic regions that require at least six, seven or eight jets,
well beyond the multiplicities required in previous analyses.
No evidence is found for physics beyond the Standard Model.
The results are interpreted in the context of a supersymmetry model (\msugra{})
where they extend previous constraints. 
\end{abstract}

\maketitle

\maybesection{Introduction} 
Many extensions of the Standard Model predict the existence of 
TeV-scale states that rapidly decay 
to a large number of strongly interacting particles in association 
with one or more stable, weakly interacting particles.
If such states are kinematically accessible with the proton-proton collisions at the LHC~\cite{Evans:2008zzb},
they will be characterised by events with
many hadronic jets with unbalanced momenta
in the plane perpendicular to the beams
due to the unobserved weakly interacting particles.

The most sensitive direct searches~\cite{daCosta:2011qk,Aad:2011ks,Aad:2011xw,ATLAS-ZERO-LEPTON-2011,Collaboration:2011ida,Chatrchyan:2011ek,*Chatrchyan:2011nd}
have been previously performed at the LHC by the
ATLAS and CMS collaborations
in selections requiring jets and
missing transverse momentum ($\ourmagptmiss$). 
These analyses required a varying number of jets from
as few as one~\cite{Chatrchyan:2011nd,Aad:2011xw}, 
to as many as $\ge$4 \cite{ATLAS-ZERO-LEPTON-2011}.

The delivery of a large ($>$\,\ifb{}) integrated luminosity 
makes it possible to extend those 
searches to final states with at least six, seven or even eight jets.
Selecting events with larger jet multiplicities provides increased
sensitivity to models that predict many-body decays or sequential 
cascade decays to many strongly interacting particles.
Such models include supersymmetric~\cite{susy_all_refs_no_titles} (SUSY) models 
that have gluinos with masses near the TeV scale and relatively heavy squarks.

Standard Model predictions must be determined with particular care
for large jet multiplicities.
The background from multi-jet events, in which the momentum imbalance
results from jet mismeasurement, is evaluated using entirely data-driven methods.
Electroweak and top contributions are obtained from a mixture of control measurements,
and transfer factors calculated from sophisticated multi-leg Monte Carlo simulations~\cite{Mangano:2002ea,Gleisberg:2008ta,*Schumann:2007mg,*Hoeche:2009rj}.
A detailed description of the background determination can be found in \oursecref{sec:backgrounds}.

Events containing high transverse momentum ($\pT$) electrons or muons are vetoed in order to reduce backgrounds 
from (semi-leptonically) decaying top quarks or $W$ bosons. 
Other complementary searches have been performed by the ATLAS collaboration in final states with 
\ourmagptmiss{} and lower jet multiplicity requirements~\cite{daCosta:2011qk,Aad:2011xw,ATLAS-ZERO-LEPTON-2011}, in conjunction with $b$-jet tagging~\cite{Aad:2011ks}, 
hard electrons and/or muons~\cite{Collaboration:2011hh,*Aad:2011xk,ATLAS-ONE-LEPTON-SUSY11} 
or photons~\cite{Aad:2011kz}.

While the results are presented in the context of \mSUGRA{}~\citemsugraandcmssm{}, the analysis is sensitive to any new states 
that decay into large numbers of jets in association with weakly
interacting particles which escape the detector unseen.

\maybesection{The ATLAS Detector and Data Samples}

The ATLAS experiment~\cite{Aad:2008zzm} is a multipurpose particle
physics detector with a forward-backward symmetric cylindrical
geometry and nearly 4$\pi$ coverage in solid angle.\footnote{
ATLAS uses a right-handed coordinate system with its origin at the nominal
interaction point in the centre of the detector and the $z$-axis along the
beam pipe. Cylindrical coordinates $(r,\phi)$ are used in the transverse
plane, $\phi$ being the azimuthal angle around the beam pipe. The pseudorapidity $\eta$ is
defined in terms of the polar angle $\theta$ by $\eta=-\ln\tan(\theta/2)$.
} The layout of the detector is
dominated by four superconducting magnet systems, which comprise a
thin solenoid surrounding inner tracking detectors 
and a barrel and two end-cap toroids supporting a large muon tracker.
The calorimeters are of particular importance to this analysis.  In the pseudorapidity
region $\left|\eta\right| < 3.2$, high-granularity liquid-argon (LAr)
electromagnetic (EM) sampling calorimeters are used.  An iron-scintillator tile
calorimeter provides hadronic coverage for 
$\left|\eta\right| < 1.7$.  The end-cap and forward regions, spanning
$1.5 < \left|\eta\right| < 4.9$, are instrumented with LAr calorimetry
for both EM and hadronic measurements.

The data sample used in this analysis was taken during the first half
of 2011 
with the LHC operating at a centre-of-mass energy of $\sqrt{s}=$7~\TeV.
Application of beam, detector and data-quality requirements resulted in
an integrated luminosity of ${1.34 \pm 0.05\,\ifb{}}$.
The analysis makes use of dedicated multi-jet triggers,
the details of which changed during the data-taking period
as a consequence of increasing LHC luminosity.
In all cases the trigger efficiency was greater than 95\% for events with 
either at least four jets with $\pt{}>80\,\GeV$, 
or at least five jets with $\pT{}>55\,\GeV$.

\maybesection{Object Reconstruction}\label{sec:objectselection}

The definitions of jets, leptons ($e$ and $\mu$)
and missing transverse momentum follow 
closely those of previous ATLAS searches~\cite{ATLAS-ZERO-LEPTON-2011,ATLAS-ONE-LEPTON-SUSY11}.

Jet candidates are reconstructed using the
anti-$k_t$ jet clustering algorithm~\cite{Cacciari:2008gp,*Cacciari:2005hq} with
distance parameter $0.4$. The inputs to this algorithm are clusters
of calorimeter cells~\cite{topoclusters} seeded by those with energy significantly above
the measured noise. Jet momenta are constructed by performing a
four-vector sum over these topological clusters of calorimeter cells, 
treating each as an
$(E,\vec{p})$ four-vector with zero mass.  These jets are corrected for
the effects of calorimeter non-compensation and inhomogeneities by
using $\ourpt$- and $\eta$-dependent calibration factors
based on Monte Carlo (MC) simulations validated with extensive
test-beam and collision-data studies~\cite{atlas-jetcal}. 
Only jet candidates with $\ourpt > 20$
\GeV{} and $|\eta| < 4.9$ are retained.
During the data-taking period, a localized electronics failure in the LAr barrel calorimeter created an electronically dead region in the second and third calorimeter layers, approximately $1.4\times 0.2$ in $\Delta\eta\times\Delta\phi$, in which on average 30\% of incident jet energy is lost. The impact on reconstruction efficiency for $\pt$ $>$ 20~GeV jets is found to be negligible.
Since the energy response for jets in the problematic region is underestimated due
to this extra dead area, a correction factor is applied to the jet transverse
momenta. Events are rejected if the correction applied to any jet candidate 
provides a contribution to $\ourmagptmiss$ that is 
greater than both $10\,\GeV$ and $0.1\,\ourmagptmiss$.
When identification of jets containing heavy flavour quarks is required, 
either to make measurements in control regions or for cross checks, 
a tagging algorithm exploiting both impact parameter and secondary vertex information is used~\cite{ATLAS-CONF-2011-102}.

Electron candidates are required to have $\ourpt$ $>$ $20~\GeV$ and $|\eta|$ $<$
2.47, to pass the `medium' electron shower shape and track selection criteria of
Ref.~\cite{Aad:2010yt}, and to be outside problematic regions of the calorimeter.
Muon candidates are required to have $\ourpt$ $>$ 10~GeV and $|\eta|$ $<$ 2.4. \footnote{When defining control regions that require the presence of one or more leptons, additional requirements are applied. Electrons must pass the `tight' selection criteria of Ref.~\cite{Aad:2010yt}, and the sum $\Sigma$ of the transverse momentum of tracks within a cone of $\Delta R=0.2$ around the electron must satisfy $\Sigma/\pT(e) < 0.1$. Muons must have longitudinal and transverse impact parameters within 1\,mm and 0.2\,mm of the primary vertex, respectively, 
and must have $\Sigma<1.8\,\GeV$.}

The measurement of the missing transverse momentum two-vector
$\ourvecptmiss$ (and its magnitude $\ourmagptmiss$) is then based on
the transverse momenta of all electron and muon candidates, all jets which are not also electron candidates with $|\eta|<4.5$, and all calorimeter clusters with $|\eta|<4.5$ not associated to such objects.

Following the steps above, overlaps between candidate jets with $|\eta|<2.8$ 
and leptons are resolved as follows.
First, any such jet candidate lying within a distance 
$\Delta R<0.2$ of an electron is discarded,
where $\Delta R = \sqrt{(\Delta \eta)^2 + (\Delta \phi)^2}$.
Then any lepton candidate remaining within a distance
$\Delta R =0.4$ of such a jet candidate is discarded.
Thereafter, all jet candidates with $|\eta|>2.8$ are discarded, 
and the remaining electron, muon and jet candidates are
retained as reconstructed objects.

\maybesection{Event Selection}\label{sec:eventselection}

Following the object reconstruction described in \oursecref{sec:objectselection}, 
events are discarded if any electrons or muons
remain, or if they contain any jet
failing quality selection criteria designed to suppress detector noise and
non-collision backgrounds
\cite{atlas-jet-clean}, or if they lack a reconstructed primary
vertex with five or more associated tracks.

\begin{table}
\small
\renewcommand\arraystretch{1.2}
 \begin{center}
   \begin{tabular}{ | l || c | c | c | c | }
     \hline
{\bf Signal region} & ~{\bf 7j55}~ & ~{\bf 8j55}~ & ~{\bf 6j80}~ & ~{\bf 7j80}~ \\ \hline
\hline
     Jet $p_{T}$ &  \multicolumn{2}{|c|}{$>$\,55\,\GeV{}} &  \multicolumn{2}{|c|}{$>$\,80\,\GeV{}}  \\ \hline
     Jet $|\eta|$ &  \multicolumn{4}{|c|}{$<$\,2.8} \\ \hline
     $\Delta R_{jj}$ &  \multicolumn{4}{|c|}{$>$\,0.6 for any pair of jets} \\ \hline
     Number of jets & $\geq$\,7 & $\geq$\,8 & $\geq$\,6 & $\geq$\,7  \\ \hline
     $\met/\sqrt{H_{T}}$ & \multicolumn{4}{|c|}{$>\,3.5\,\rootgev$}   \\ \hline
   \end{tabular}
   \caption{\label{tab:sr}
     Definitions of the four signal regions.
   }
 \end{center}
\end{table}

Four different signal regions (SRs) are defined as shown in
\ourtabref{tab:sr}. The use of multiple signal regions provides sensitivity
in different areas of the \msugra{} plane. Furthermore, the
complementarity of the selections may be enhanced in new models not
explicitly considered here.
The combinations of jet multiplicities and $\pT$ thresholds
are chosen such that all four SRs have trigger efficiencies in excess of $95\%$ and
acceptances greater than 2\%~-~3\% for kinematically accessible \msugra{} models.
Differences caused by jet merging and splitting between the offline and online selections 
can lead to trigger inefficiencies.
A separation of $\Delta R_{jj}>0.6$ between all jets with \pT{} above the threshold
for the SR is required to maintain acceptable trigger efficiency. 

The final selection variable is $\metsig$, the ratio 
of magnitude of the missing transverse momentum
to the square root of the scalar sum \HT{} of transverse momenta of all jets with $\pT>40\,\GeV$ and $|\eta|<2.8$.
This ratio provides a measure of the size of the missing transverse momentum
relative to the resolution due to stochastic variations in the measured jet energies.

\maybesection{Backgrounds, Simulation and Normalisation}\label{sec:backgrounds}

Standard Model processes contribute to the event counts in the signal regions.
The dominant backgrounds are multi-jet production, 
including those from purely strong interaction processes and
fully hadronic decays of \ttbar{};
semi- and fully-leptonic decays of \ttbar{};
and leptonically-decaying $W$ or $Z$ bosons produced in association with jets.
Non-fully-hadronic top, and $W$ and $Z$ are collectively 
referred to as \leptonic{} backgrounds, and can contribute
to the signal regions when no $e$ or $\mu$ leptons are produced 
(for example $Z\rightarrow \nu\nu$ or hadronic $W\rightarrow\tau\nu$ decays)
or when they are produced but out of acceptance or not reconstructed.
Contributions from the hadronic decays of $W$ and $Z$ bosons are negligible.

The selection cuts were chosen such that 
the background from the multi-jet processes can be determined from supporting measurements.
In events dominated by jet activity,
the ATLAS \ourmagptmiss{} resolution is approximately proportional to $\sqrt{\HT}$. 
The ratio $\metsig$ is therefore almost invariant under changes in the
jet multiplicity \njet{}, as will be shown later.

The shape of the \metsig{} distribution for the multi-jet background 
is therefore determined from data using control regions 
CR with smaller jet multiplicities than the SRs.
The control regions are assumed to be dominated by Standard Model processes, an assumption that is 
corroborated by the agreement of multi-jet cross section measurements with up to six jets~\cite{Collaboration:2011tq} with Standard Model predictions.
The signal `contamination' contributes less than $1\%$  
to the higher multiplicity CRs for relevant, unexcluded \mSUGRA{} points.
The basic shape of the \metsig{} distribution is
encapsulated in transfer factors $T_\textrm{j,p}$, defined to be the ratio of the number of events 
in the control region, $\mathrm{CR}_\textrm{j,p}^+$, with a certain number, $j$, of jets above a $\pT\,$ threshold $p$ and $\metsig>3.5\rootgev$ 
to the number in the control region, $\mathrm{CR}_\textrm{j,p}^-$, with the same $j,p$ and $\metsig<1.5\,\rootgev$.
The $T_\textrm{j,p}$ are calculated after subtracting the 
predicted contributions of the \leptonic{} backgrounds
from the measured counts.
The signal region prediction is found
by applying a $T_\textrm{j,p}$, with the same $p$ as the signal region and $j=5$ when $p=55\,$GeV and $j=4$ when $p=80\,$GeV, to the number of events (after subtracting the expected contribution
from \leptonic{} background sources) satisfying signal region multiplicity requirements but with $\metsig<1.5\,\rootgev$.

\begin{figure*}
\begin{center}
\subfigure[]{\includegraphics[width=\plotwidth]{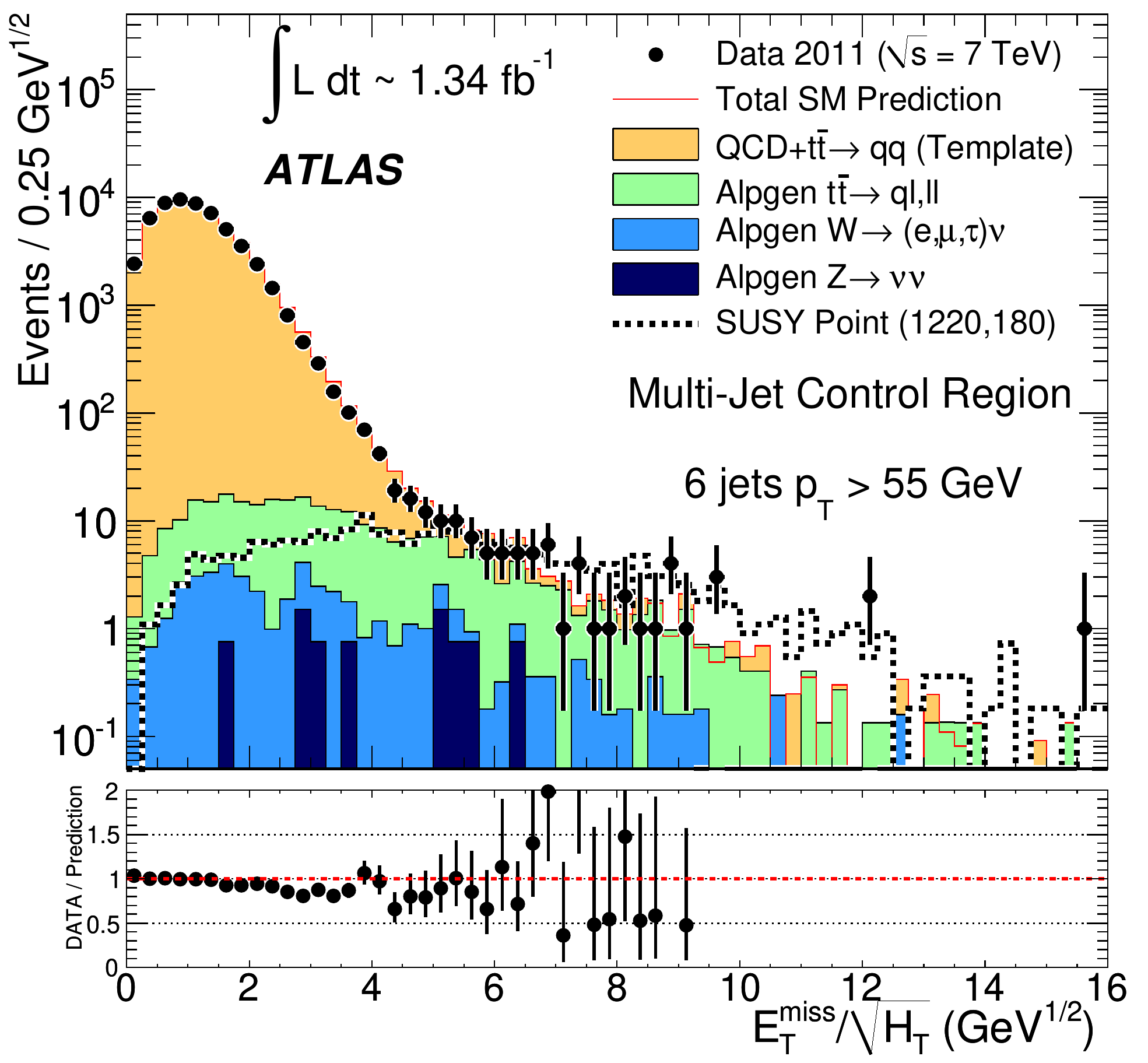}\label{h5j30_MetSig6}}
\subfigure[]{\includegraphics[width=\plotwidth]{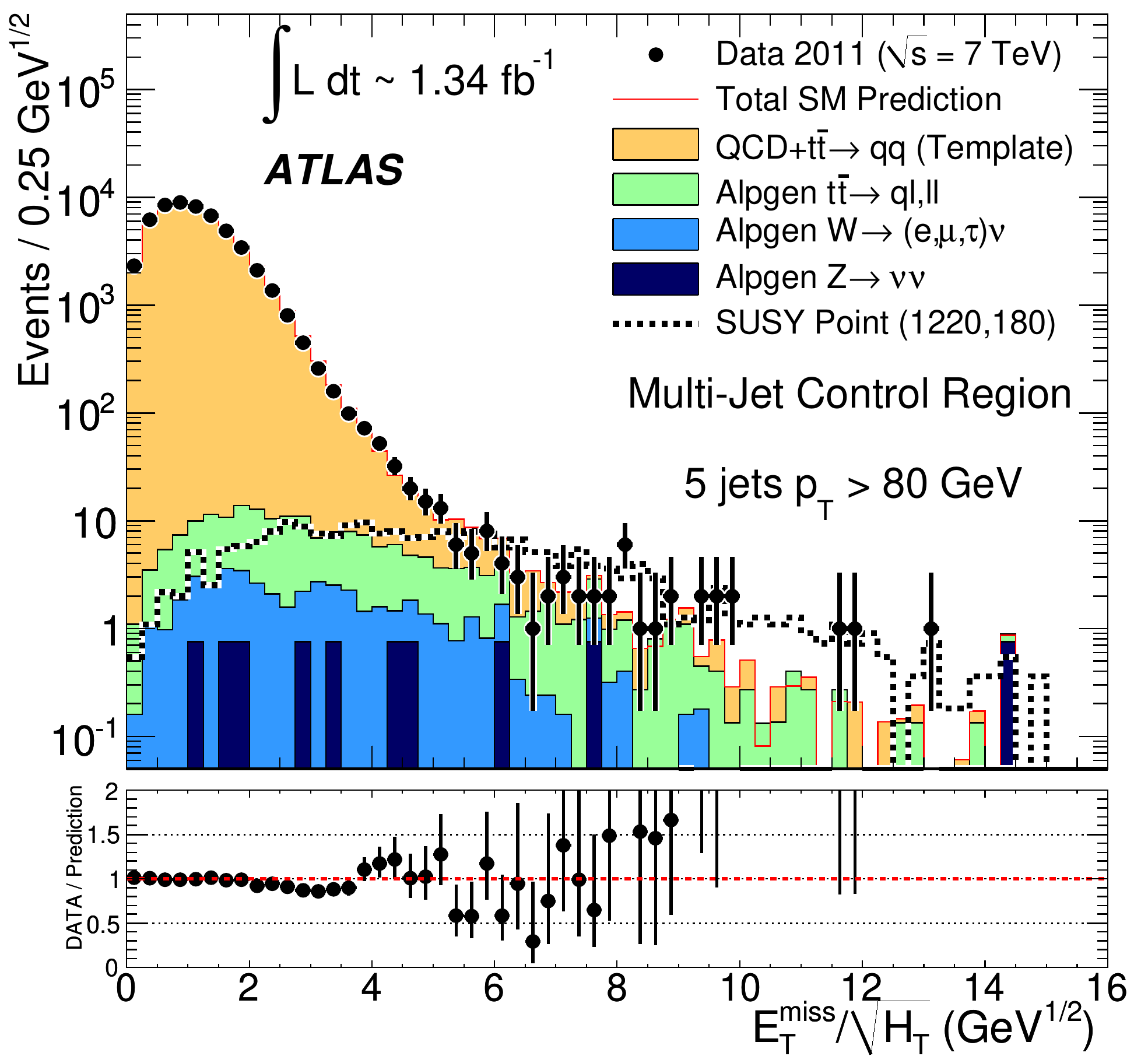}\label{h4j45_MetSig5}}
\caption{\label{fig:metsig_bg}
  The distribution of the variable $\met{}/\sqrt{H_T}\,$
  for control regions requiring \oursubref{h5j30_MetSig6} exactly six jets with $\pT>55\,\GeV$
  or \oursubref{h4j45_MetSig5} exactly five jets with $\pT>80\,\GeV$. 
  Overlaid are templates taken from selections requiring \oursubref{h5j30_MetSig6} exactly five jets
  with $\pT>55\,\GeV$ or \oursubref{h4j45_MetSig5} exactly four jets with $\pT>80\,\GeV$. These templates are normalised 
  to the data in the region with $\met{}/\sqrt{H_T}<1.5$.
  The background estimation includes the \alpgen{} Monte-Carlo prediction 
  for the \leptonic{} Standard Model backgrounds.
  For illustrative purposes the plots also contain the distribution expected for an example \mSUGRA{} 
  point with $m_0=1220$\,\GeV{} and $m_{1/2}=180$\,\GeV{}.
  In all ratio plots (lower), in regions with very low numbers of events, some points may lie off the shown range and for bins with no observed events no ratio is shown.  
  For $\ttbar$ backgrounds the labels `qq' and `ql,ll' represent fully hadronic and non-fully-hadronic decays respectively. 
}
\end{center}
\end{figure*}

The validity of the assumption of \metsig{} invariance has been tested
with data, using a series of additional CRs 
with either smaller jet multiplicities than the SRs,
or at smaller values of \metsig{} (between 1.5\,\rootgev and 3.0\,\rootgev) or both.
Templates are formed from data selections with lower values of \njet{} and correcting for \leptonic{}
contamination. After scaling by the appropriate normalisation
the shapes of the \metsig{} distributions in these CRs
are found to be well described by these templates
(see Figures~\ref{fig:metsig_bg} and \ref{fig:vrmult} for examples).
The numbers of events in each of six different CRs are found to be correctly predicted
to within $\sim 10\%-20\%$. The residual differences
are included in the systematic uncertainty associated with the method.

\begin{figure*}
\begin{center}
\subfigure{\includegraphics[width=\plotwidth]{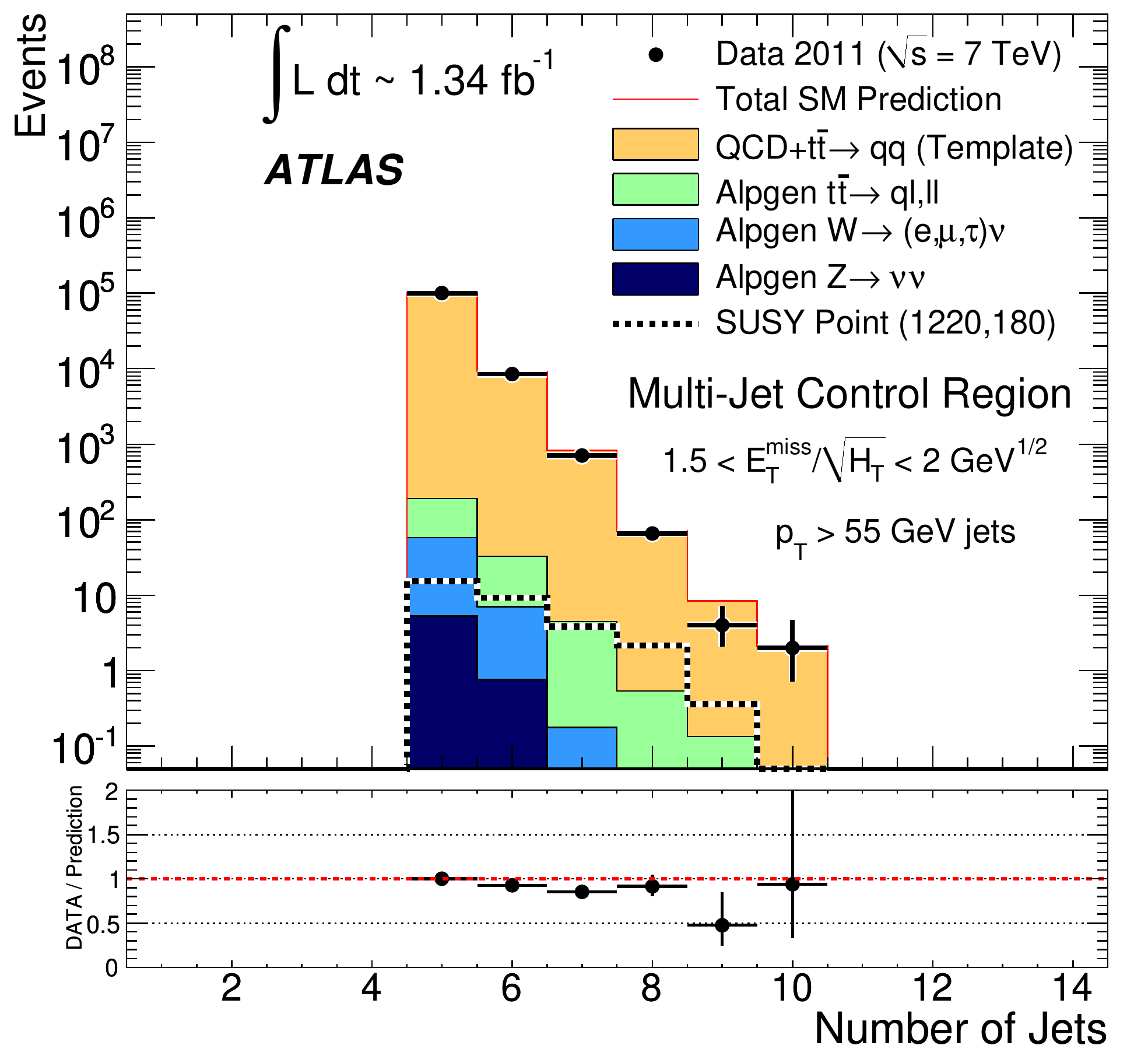}\label{mult_55_15_20}}
\subfigure{\includegraphics[width=\plotwidth]{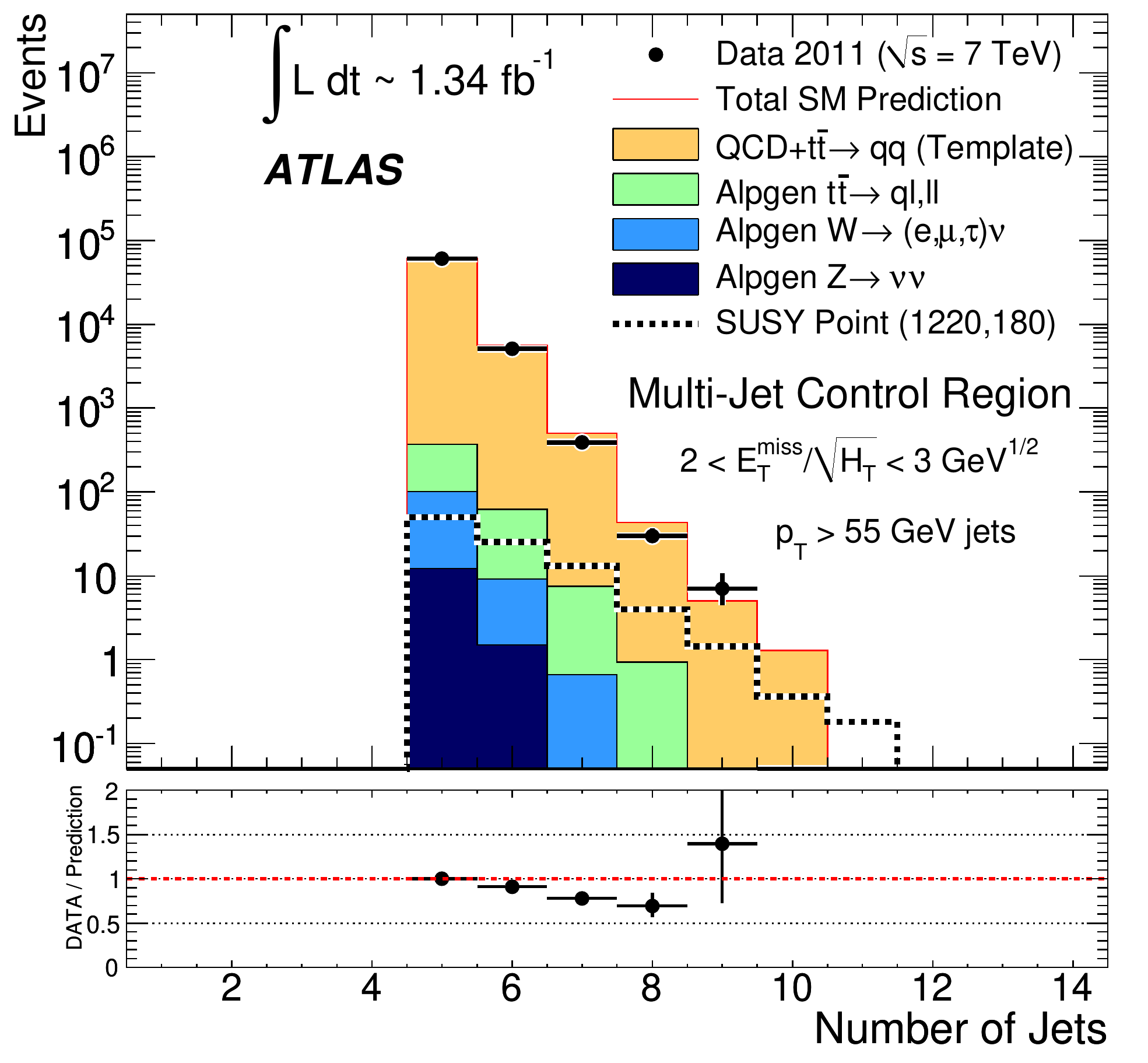}\label{mult_55_20_30}}\\
\subfigure{\includegraphics[width=\plotwidth]{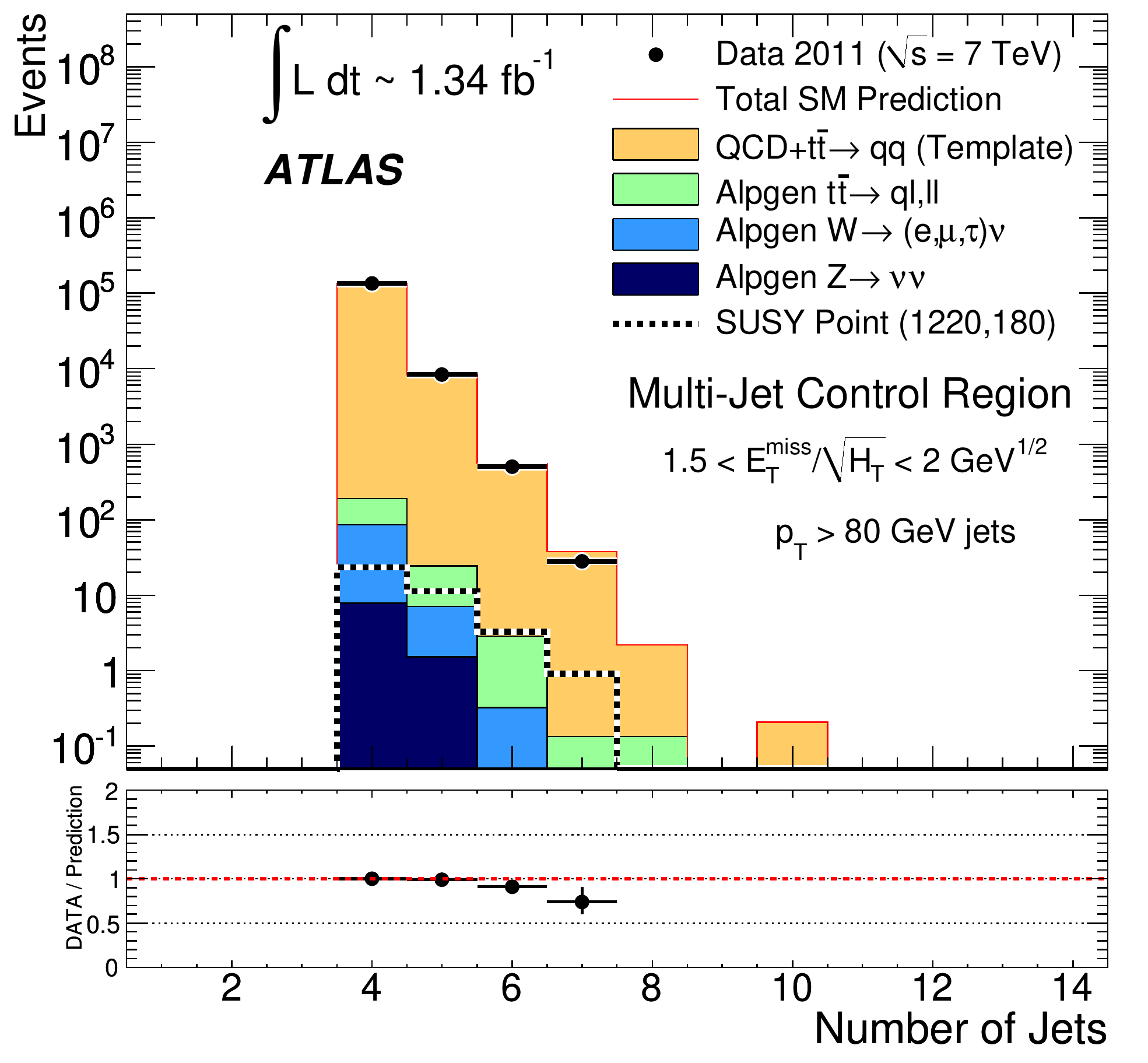}\label{mult_80_15_20}}
\subfigure{\includegraphics[width=\plotwidth]{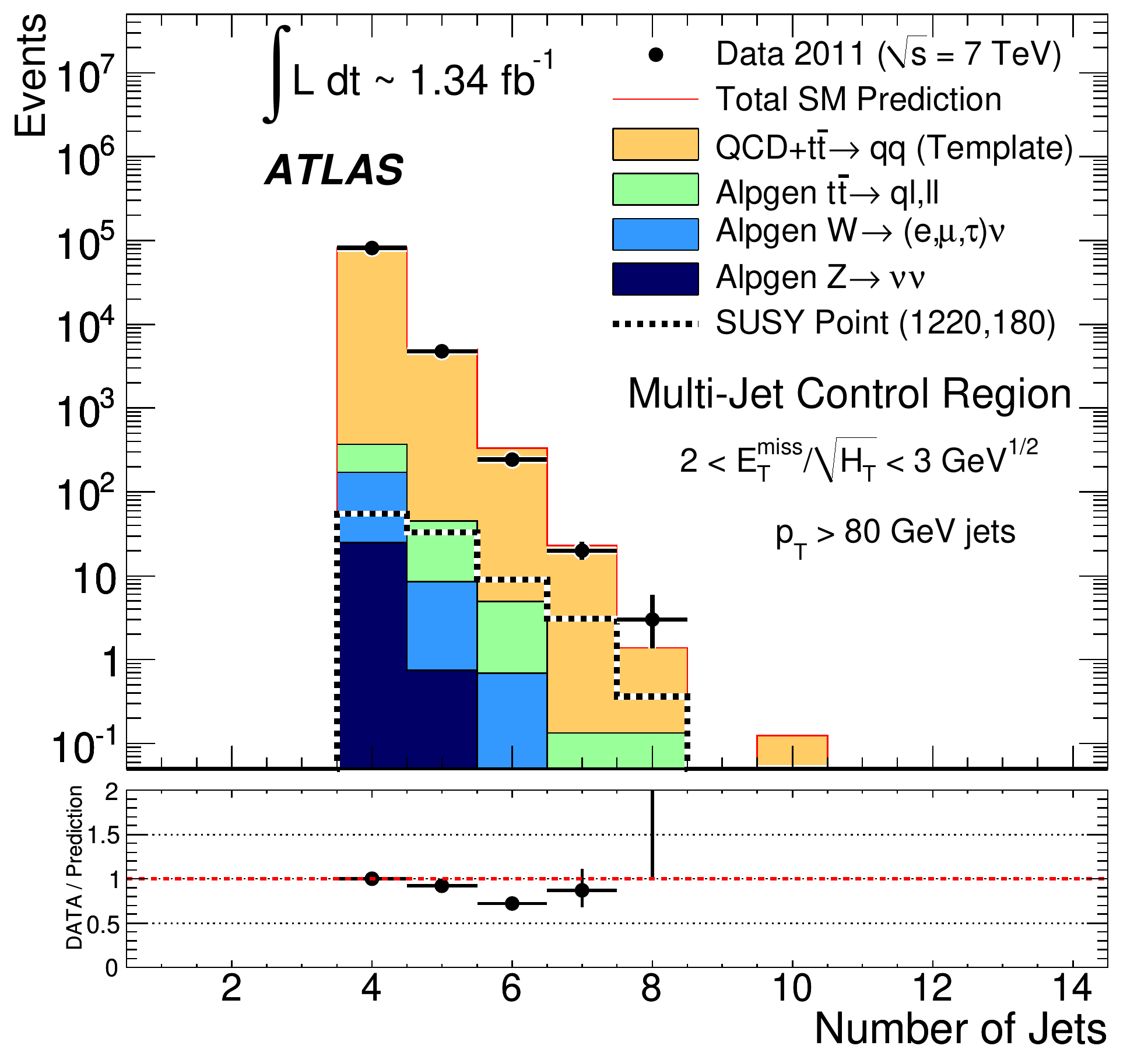}\label{mult_80_20_30}}
\end{center}
\caption{\label{fig:vrmult}
Observed and predicted jet multiplicity distributions for jets with $\pT{}>55$~GeV (upper)
and with $\pT{}>80$~GeV (lower)
in four example control regions defined by
$1.5\,\rootgev<\metsig<2\,\rootgev$ (left) and $2\,\rootgev<\metsig<3\,\rootgev$ (right). 
Overlaid are templates taken from selections requiring $\metsig<1.5\,\rootgev\,$ which are normalised to the 
data in the lowest jet multiplicity bin shown.
The background estimation includes the \alpgen{} Monte-Carlo prediction 
for the \leptonic{} Standard Model backgrounds.
For illustrative purposes the plots also contain the distribution expected for an example \mSUGRA{} 
point with $m_0=1220$\,\GeV{} and $m_{1/2}=180$\,\GeV{}.
}
\end{figure*}

The backgrounds from multi-jet processes are cross checked using another 
data-driven technique~\cite{daCosta:2011qk,ATLAS-ZERO-LEPTON-2011} 
which smears the energies of individual jets from low-\met{} multi-jet `seed' events in data.
Separate smearing functions are defined for $b$-tagged and non-$b$-tagged
jets, with each modelling both the Gaussian core and the non-Gaussian tail of the jet response.
The functions are based on simulations and are 
verified with data in three-jet control regions in which the \ourvecptmiss{} can be associated 
with the fluctuation of a particular jet.
The agreement between the two methods is satisfactory within uncertainties, and 
so the prediction used in what follows is that based on \metsig{} shape invariance.

Monte Carlo simulations are used 
to determine the transfer factors used to estimate the \leptonic{}
Standard Model backgrounds, 
and to assess sensitivity to specific SUSY signal models.
When used for \leptonic{} background estimation, the resulting transfer factors
connect CRs and SRs with similar 
selection requirements. 
Theoretical uncertainties, including those arising from
the use of Leading Order (LO) generators, are therefore reduced.
All Monte Carlo samples employ a GEANT4~\cite{Agostinelli:2002hh} based 
detector simulation~\cite{:2010wqa}, and are reconstructed with the same algorithms as the data.
The simulations include the effects of multiple proton-proton interactions
per bunch crossing.

\begin{figure*}
\centering
\subfigure[]{\includegraphics[width=\plotwidth]{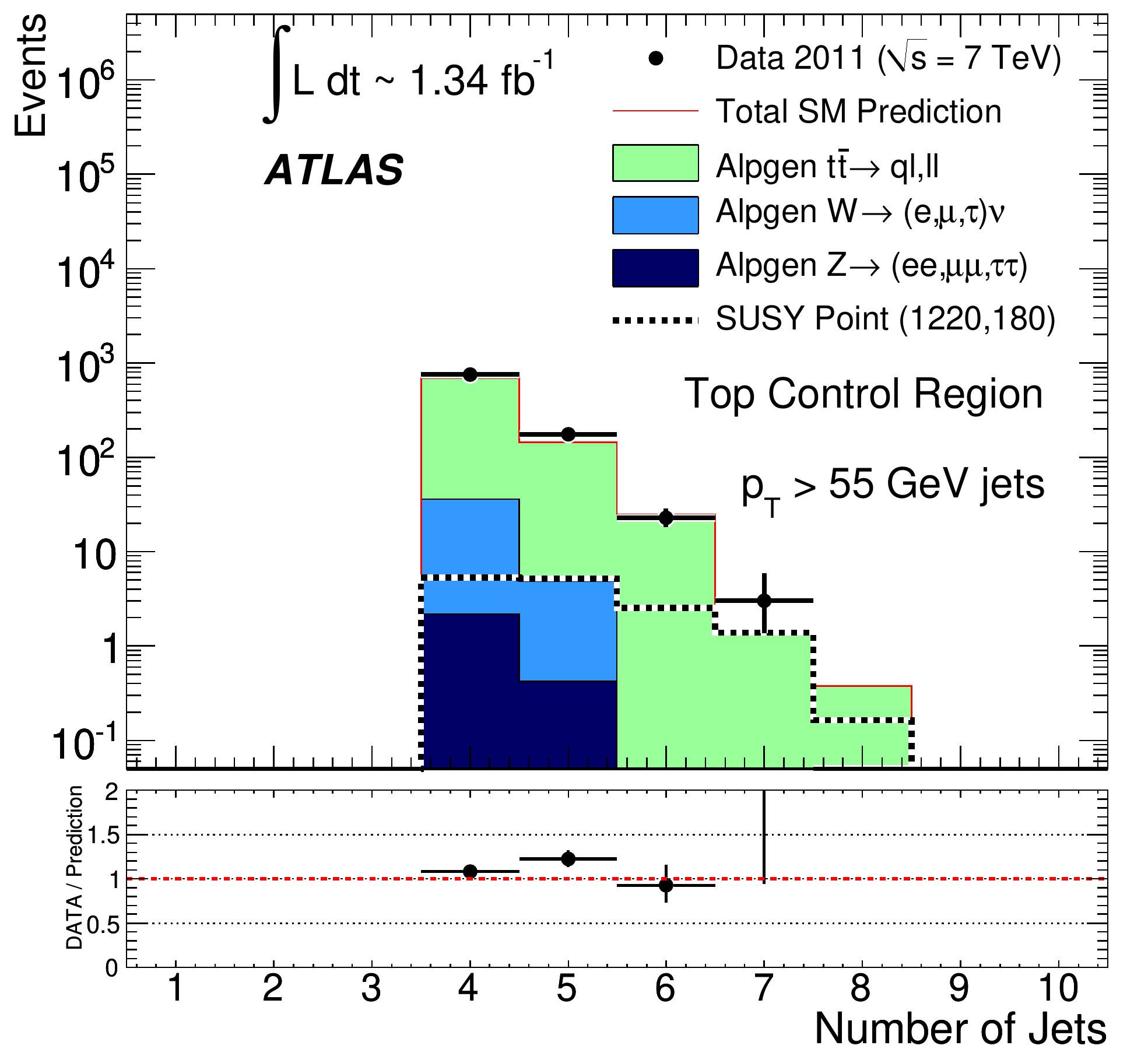}\label{fig:ttbar_cr_55}}
\subfigure[]{\includegraphics[width=\plotwidth]{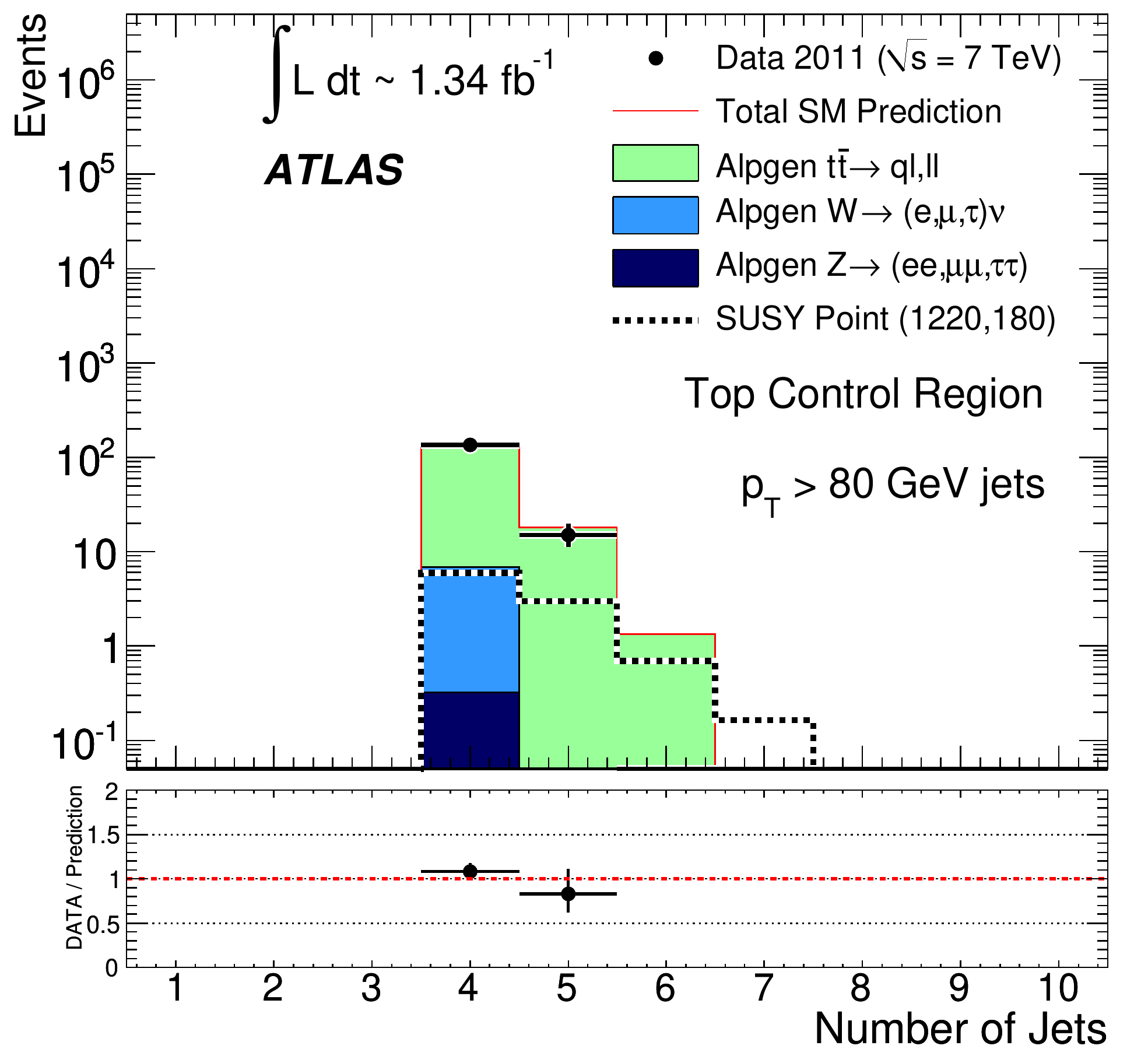}\label{fig:ttbar_cr_80}}
\\
\subfigure[]{\includegraphics[width=\plotwidth]{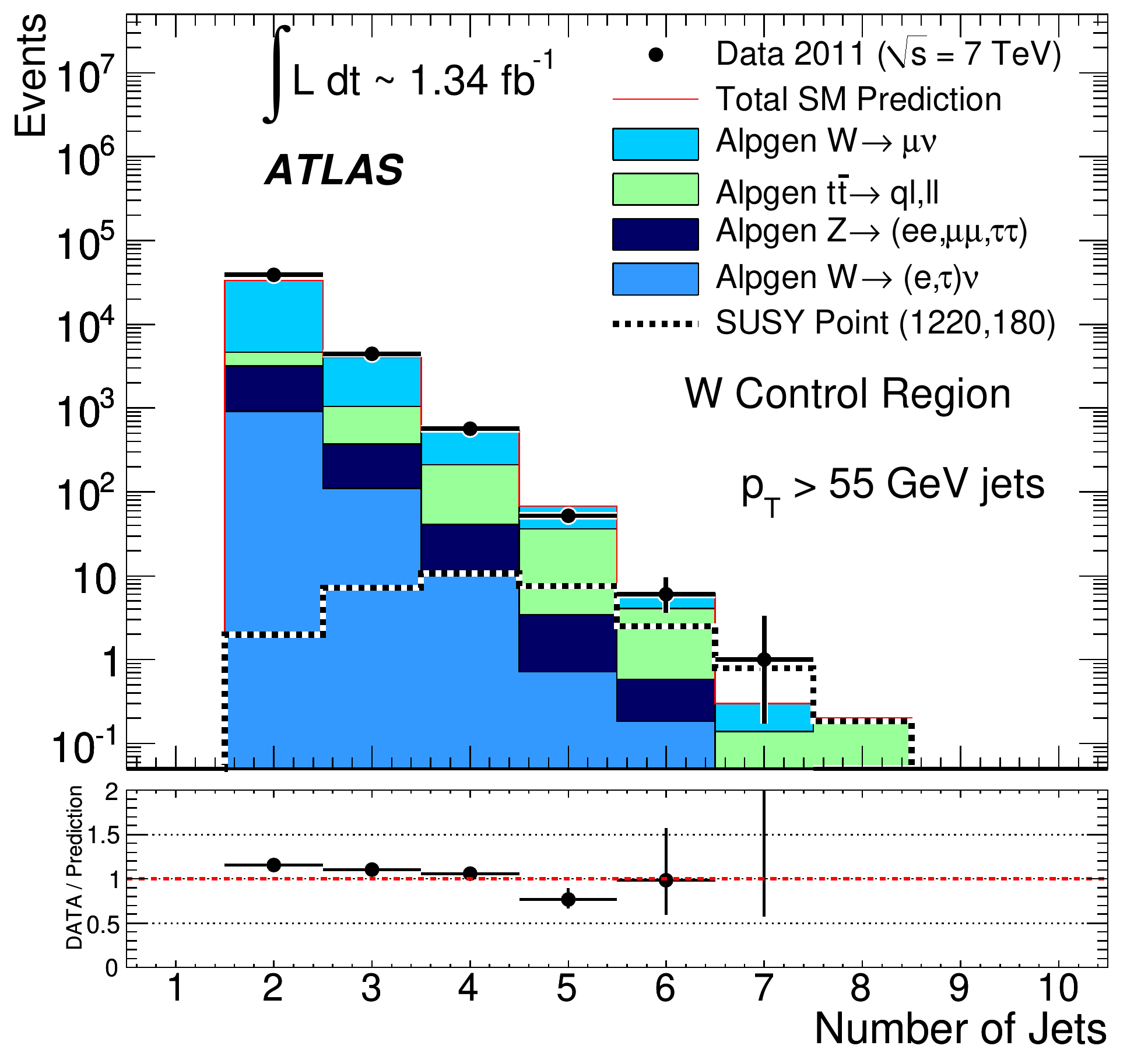}\label{fig:w_mult_55}}
\subfigure[]{\includegraphics[width=\plotwidth]{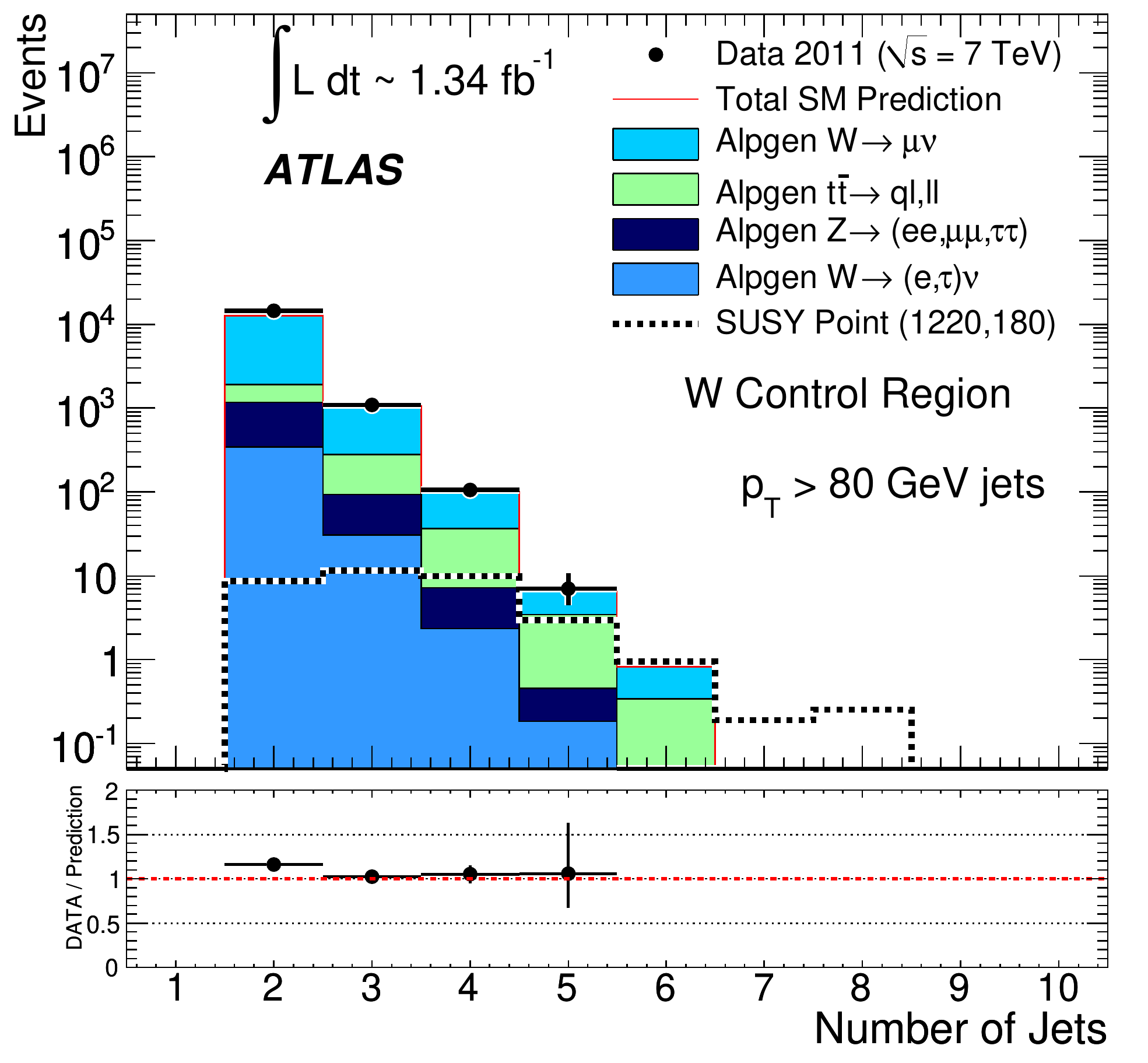}\label{fig:w_mult_80}}
\\
\subfigure[]{\includegraphics[width=\plotwidth]{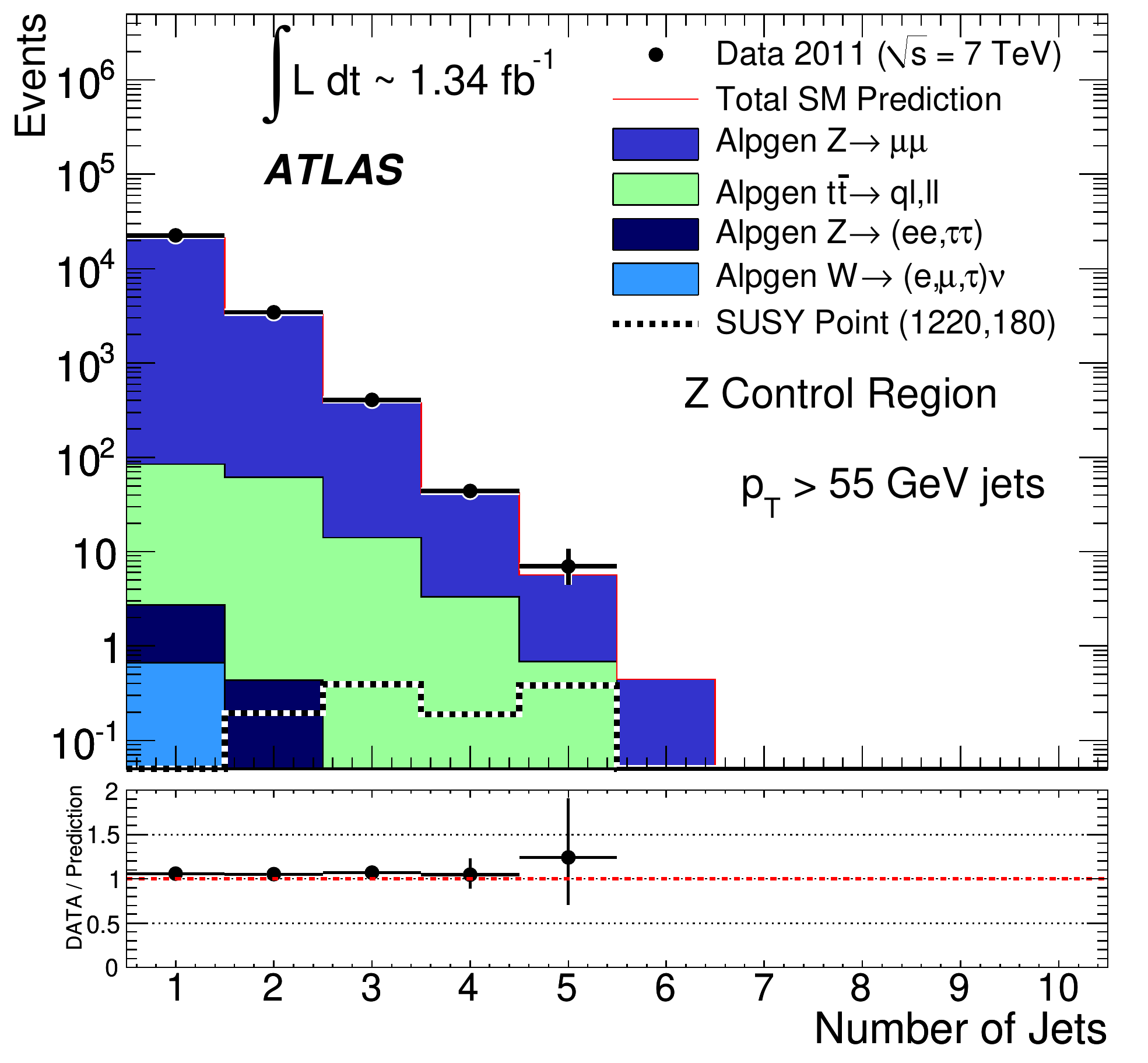}\label{fig:z_mult_55}}
\subfigure[]{\includegraphics[width=\plotwidth]{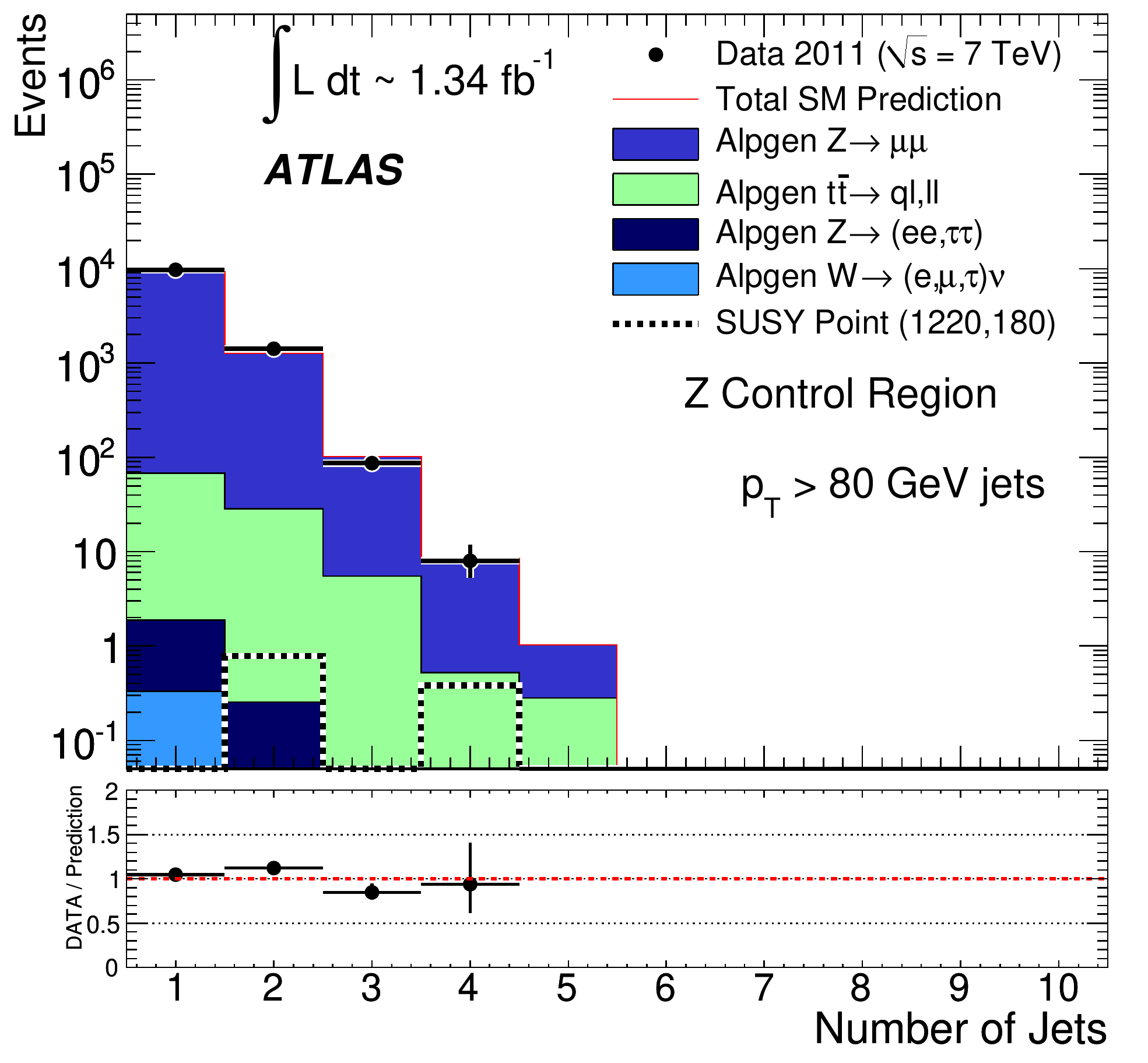}\label{fig:z_mult_80}}
\caption{The multiplicity of jets with $\pT>55\,\GeV{}$ (left) or $\pT>80\,\GeV{}$ (right)
for events in various control regions.
Top row: top-quark enhanced control regions requiring at least one $b$-tagged jet, a single isolated muon with $p_T>20\,$GeV and $|\eta|<2.4$ and $40<\mT<100\,$GeV.
Middle row: $W$-boson enhanced region (as for top, but with $b$-jet veto). 
Bottom row: $Z\rightarrow \mu\mu$ enhanced region. 
}
\label{fig:ttbar_cr}
\end{figure*}

To estimate the contribution from \leptonic{} \ttbar{} events, 
control regions are defined 
with exactly one $\pT>20\GeV$ muon, transverse mass\footnote{The transverse mass is defined by $\mT^2(a,b)=m_a^2 + m_b^2 + 2 \left(E_{\rm T}^{(a)}E_{\rm T}^{(b)}-{\vec p}_{\rm T}^{\,(a)}\cdot {\vec p}_{\rm T}^{\,(b)}\right)$, where $E_T^2=\pT^2 + m^2$. The massless representation is used for $\ourvecptmiss$.}
in the range $40\,\GeV<\mT(\mu,\ourvecptmiss)<100\GeV$, 
and at least one $b$-tagged jet.
The jet multiplicity distributions for this initial control selection 
are shown in Figures~\ref{fig:ttbar_cr_55} and \ref{fig:ttbar_cr_80}.
The control regions, \topcr{}, are then formed by including the leptons 
in the jet multiplicity, $j$, and requiring $\metsig>3.5\,\rootgev$. 
The contribution in each SR is calculated from the corresponding \topcr{} (with the same $j,p$)
in each case using a transfer factor evaluated using \alpgen{}~\cite{Mangano:2002ea}~v2.13 
\ttbar{} Monte Carlo, the PDF set {\tt CTEQ6L1}~\cite{Pumplin:2002vw} and up to three (and as a cross check, up to five)
additional outgoing partons in the matrix elements.
Parton showering, fragmentation and hadronization for all \alpgen{} samples is
performed with \herwig~\cite{Corcella:2000bw,*herwig65long}, while
\jimmy~\cite{Butterworth:1996zw} is used to simulate the underlying event. 

The vector boson processes $Z\rightarrow \nu\nu$ and $W\rightarrow \ell\nu$
provide small contributions to the signal regions when produced in association 
with jets. 
The $W\rightarrow\ell\nu$ + jets background is evaluated using 
an \alpgen{}-based simulation with up to five additional partons in the matrix elements.
Control regions are defined with selections similar to the \topcr{}  
but with a $b$-jet veto to reduce contamination from \ttbar{}.
The jet multiplicity distributions for the $W$-enhanced selection 
can be found in Figures~\ref{fig:w_mult_55} and \ref{fig:w_mult_80}.

The $Z\rightarrow \nu\nu$ + jets background is calculated using an \alpgen{}-based
simulation with up to five additional partons in the matrix elements.
There are sufficient $Z\rightarrow \mu^+\,\mu^-$ events in data to 
verify the Monte Carlo predictions for multiplicities in the range $1 \leq \njet\leq 5$~(Figures~\ref{fig:z_mult_55} and \ref{fig:z_mult_80}).
The ratio of cross sections $R_n \equiv \frac{\sigma \left(V + (n+1)\,{\rm jets}\right)} {\sigma \left(V + n\,{\rm jets}\right)}$ is found, both in simulations and in data,
to be nearly constant.\footnote{
See also Ref~.\cite{Ellis:1985vn,*Berends:1989cf,*Berger:2010gf,*Abazov:2002gy,*Wagner:2003cv,*:2010pg}.}
The values of $R_n$ for $\njet\in\{5,\,6,\,7\}$ have been cross checked against {\tt SHERPA}~\cite{Gleisberg:2008ta,*Schumann:2007mg,*Hoeche:2009rj} with {\tt COMIX}~\cite{Gleisberg:2008fv} and agreement found at the 20\% level.

Backgrounds from other sources such as
single top or diboson production, which have been evaluated using Monte Carlo simulations,
and those from non-collision sources are found to be negligible.

\msugra{} particle spectra and decay modes are calculated with {\tt ISAJET}~\cite{isajet}~v7.75. Samples are generated with {\tt HERWIG++}~\cite{Bahr:2008pv}~v2.4.2. The cross sections are normalised using the next-to-leading-order predictions of \prospino{}~\cite{Beenakker:1996ch}~v2.1. 

\maybesection{Systematic Uncertainties}\label{sec:sysematics}

Systematic uncertainties arise through the 
imperfect modelling of the multi-jet \metsig{} distribution,
the use of MC-derived transfer factors relating observations in the control regions to \leptonic{} background expectations in the signal regions, 
and from the calculation of the SUSY signal.

For the multi-jet contribution, systematic uncertainties are determined 
to account for the residual dependence of the \metsig{} distribution on 
\njet{} (as described in \oursecref{sec:backgrounds}),
the fraction of jets containing $b$ quarks and the response in problematic areas
of the calorimeter. A special study performed to quantify the
effect of the dead area of the calorimeter found that 
after applying the correction-based veto 
described in \oursecref{sec:objectselection}
the uncertainty is less than 5\%.

Jets containing heavy flavour ($b$ and $c$) quarks
can include neutrinos and hence have broader resolution functions.
The size of the systematic uncertainty resulting from heavy flavour ($b$-jet)
contributions, including those from fully-hadronic \ttbar{},
is determined as follows. 
Separate values of $T$ are calculated for events with 
at least one $b$ jet and for non-$b$-tagged sub-samples,
and their individual contributions to the SRs are determined. 
The differences with respect to the flavour-blind determination
actually used 
are $8\%-15\%$ depending on SR, 
and are included as a systematic uncertainty.

The transfer factors calculated for the \leptonic{} backgrounds 
have systematic uncertainties due to 
the finite number of Monte Carlo events generated,
the jet energy scale, the jet energy resolution, 
the lepton identification efficiency, the $b$-tag efficiency, 
and the effect of multiple proton-proton interactions per bunch crossing.

Theoretical uncertainties on the SUSY signal were estimated from variation
of the factorisation and renormalisation scales in \prospino{} between half
and twice the mean outgoing sparticle mass 
and by considering the PDF uncertainties provided by \cteqVER{6.6}~\cite{Nadolsky:2008zw}.
Uncertainties were calculated for individual
production processes (e.g.~$\squark\squark$, $\gluino\gluino$, etc.)
and are typically $30\%-40\%$ for models in the vicinity of the limits expected to be set by this analysis. 
For the signal samples, the combined experimental systematic uncertainties from jet energy scale, resolution, and cleaning are typically $15\%-20\%$. 
The 3.7\% luminosity uncertainty~\cite{lumi_summer_11,*Aad:2011dr} is included but is negligible.

\maybesection{Results, Interpretation and Limits}\label{sec:results}

\begin{figure*}
\centering
\subfigure[]{\includegraphics[width=\plotwidth]{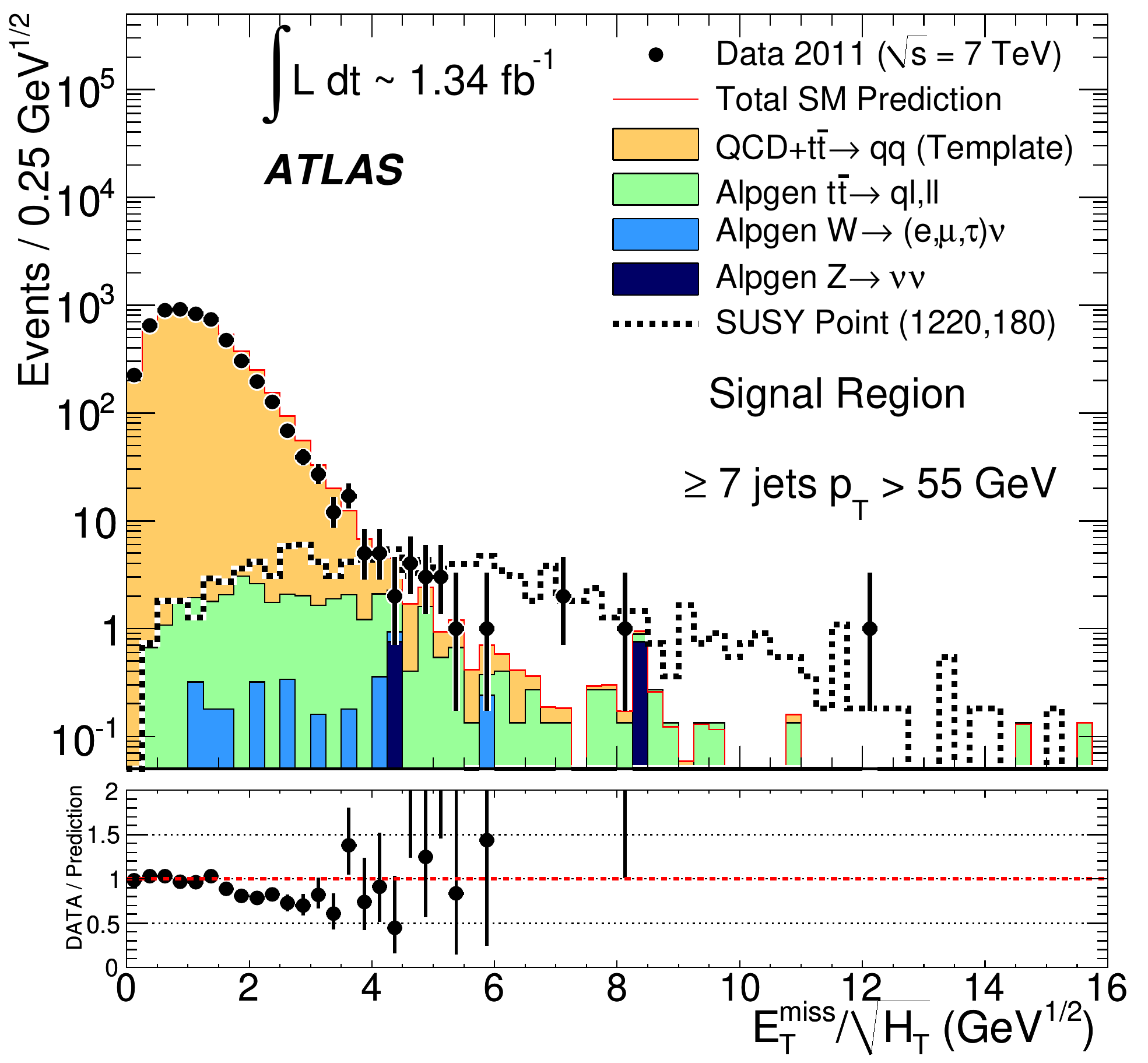}\label{fig:metsig:7j55}}
\subfigure[]{\includegraphics[width=\plotwidth]{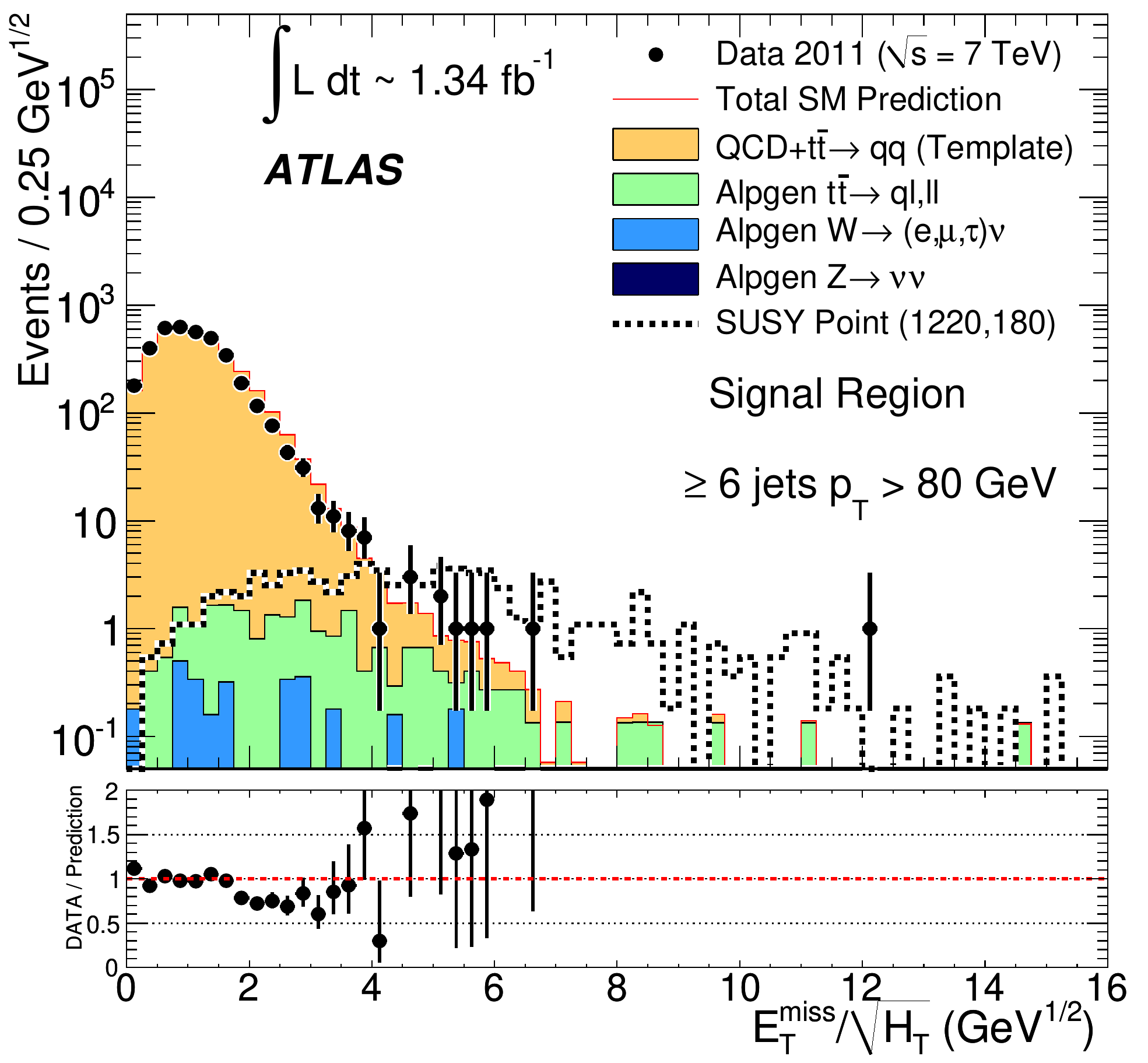}\label{fig:metsig:6j80}}
\caption{\label{fig:metsig_signal}
  The distribution of the variable $\met{}/\sqrt{H_T}\,$
  for events with
  \oursubref{fig:metsig:7j55} seven or more jets with $\pT>55\,\GeV$ or 
  \oursubref{fig:metsig:6j80} six or more jets with $\pT>80\,\GeV$. 
  Overlaid are the hadronic background templates plus 
  the \alpgen{} Monte-Carlo prediction 
  for the \leptonic{} Standard Model backgrounds.
  For illustrative purposes the plots also contain the distribution expected for an example \mSUGRA{} 
  point with $m_0=1220$\,\GeV{} and $m_{1/2}=180$\,\GeV.}
\end{figure*}

The measured \metsig{} distributions for two of the signal regions 
are shown in Figure~\ref{fig:metsig_signal} prior to the final $\metsig{}>3.5\,\rootgev$ requirement.
The number of observed events for each of the signal regions 
is shown in \ourtabref{tab:results}.
The Standard Model expectations are also shown, together with
their combined statistical and systematic uncertainties.
The data are found to be in good agreement with the background model and no excess is observed.
Table~\ref{tab:results} shows the 95\% confidence level upper bound \nmaxbsm{} on the number of events originating from 
sources other than the Standard Model, the corresponding upper limit $\sigmamaxbsm{}\times\epsilon$ on the 
cross section times efficiency within acceptance
(which equals the limit on the observed number of signal events divided by the luminosity) 
and the $p$-value for the Standard Model-only hypothesis ($p_{\rm SM}$).

An interpretation of these results is presented in
\ourfigref{fig:sugraExclusion} as a 95\% confidence level exclusion
region in the $\tan\beta=10$, $A_0=0$, $\mu>0$ slice of
\mSUGRA\footnote{A particular \mSUGRA{} model point is specified by five parameters: the universal scalar mass $m_0$, the universal gaugino mass $m_{1/2}$, the universal trilinear scalar coupling $A_0$, the ratio of the vacuum expectation values of the two Higgs fields $\tan\beta$, and the sign of the higgsino mass parameter $\mu$.}.
Data from the four SRs are used to set the limits, taking the SR with
the best expected limit at each point in parameter space. The limit
for each SR is obtained by comparing the observed event count with
that expected from Standard Model background plus SUSY signal processes, taking into account uncertainties in the expectation including those which are correlated between signal and background (for instance jet energy scale uncertainties). The exclusion regions are obtained using the CL$_{\rm s}$ prescription~\cite{clsread}.
Acceptance times efficiency values are tabulated for typical points elsewhere~\cite{multijet_2011_auxiliary}.

\begin{table}
\renewcommand\arraystretch{1.25}
\footnotesize
\begin{center}
\begin{tabular}{ | l || c | c | c | c |  } 
\hline
{\bf Signal region} & {\bf ~7j55~} & {\bf ~8j55~} & {\bf ~6j80~} & {\bf ~7j80~} \\ \hline\hline
Multi-jets &  $26\pm5.2$ & $2.3\pm0.7$ & $19\pm4$ & $1.3\pm0.4$ \\ \hline
$\ttbar{} \rightarrow q\ell,\ell\ell$ &
                   $10.8 \pm 6.7$ & $0^{+4.3}$ & $6.0 \pm 4.6$ & $0^{+0.13}$  \\ \hline
$W$ + jets    & $0.95\pm0.45$ & $0^{+0.13}$ & $0.34\pm0.24$  &  $0^{+0.13}$ \\ \hline
$Z$ + jets    & $1.5^{+1.8}_{-1.5}$ & $0^{+0.75}$ & $0^{+0.75}$ & $0^{+0.75}$ \\ \hline\hline
{\bf Total Standard Model} 
              & $39\pm9$ & $2.3^{+4.4}_{-0.7}$ & $26\pm6$  &  $1.3^{+0.9}_{-0.4}$   \\ \hline\hline
{\bf Data  }  &   45   &  4     &   26   &   3    \\ \hline\hline
\nmaxbsm{}
              &   26.0     &   11.2     &  16.3      &  6.0      \\ \hline
$\sigmamaxbsm \times\epsilon / {\rm fb}$
              &  19.4      &   8.4     &   12.2     &   4.5     \\ \hline
$p_{\rm SM}$
              &  0.30      &  0.36      &  0.49      &  0.16      \\ \hline
\end{tabular}
\caption{\label{tab:results}
Results for each of the four signal regions for \ourintlumi{}.
The expected number of Standard Model events are given for
each of the following sources:
multi-jet (including fully hadronic \ttbar{}),
semi- and fully-leptonic top combined,
and $W$ and $Z$ bosons (separately) in association with jets, 
as well as the total Standard Model expectation.
Where small event counts in control regions have not made it possible to determine 
a central value for the expectation, an asymmetric bound is given instead.
The number of observed events is also shown.
The final three rows show the statistical quantities described in the text.
}
\end{center}
\end{table}

\begin{figure}[tbh]
\begin{center}
\includegraphics[width=0.99\columnwidth]{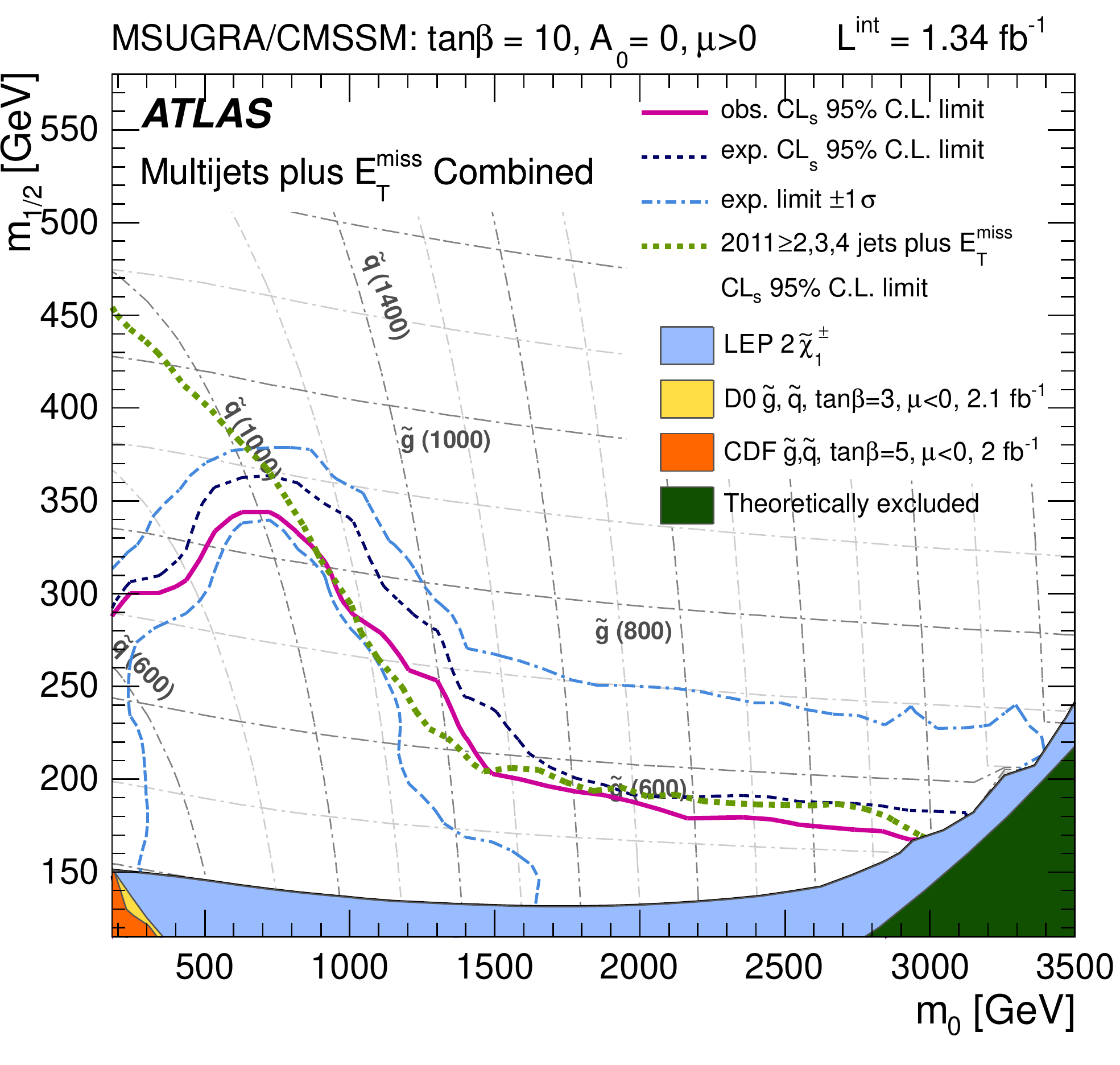}
\caption{\label{fig:sugraExclusion}
Combined exclusion bounds in the $\tan\beta=10, A_0=0, \mu>0$
slice of the \msugra{} space. Gluinos with masses below 520\,GeV,
and gluinos with masses below 680\,GeV under the assumption that
$m_{\rm squark} = 2 \times m_{\rm gluino}$ are excluded at the 95\%
confidence level.
Limits from individual SRs can be found elsewhere~\cite{multijet_2011_auxiliary}.
Recent limits from ATLAS~\cite{ATLAS-ZERO-LEPTON-2011},
as well as previous limits from D0 and CDF~\cite{Abachi:1995ng,*Affolder:2001tc,*Aaltonen:2008rv,*:2007ww}
and LEP~\cite{lepsusy} are also shown.
}
\end{center}
\end{figure}

\maybesection{Summary}

A search for new phenomena has been performed using events containing missing transverse momentum
and much larger jet multiplicities than have been previously considered, up to eight or more jets.
The dominant Standard Model background contributions have been determined from the data
themselves. The sub-dominant \leptonic{} backgrounds are measured using multiple control regions 
together with Monte Carlo transfer factors.

No evidence for physics beyond the Standard Model has been
observed in a data sample from early 2011 corresponding to
an integrated luminosity of $\ourintlumi$.
Limits are set on \mSUGRA{} models 
excluding at the 95\% confidence level gluinos with masses below 520\,GeV,
and gluinos with masses below 680\,GeV under the assumption that $m_{\rm squark} = 2 \times m_{\rm gluino}$.
This result extends those set by previous ATLAS analyses.

\maybesection{Acknowledgments}

We wish to thank CERN for the efficient commissioning and operation of the LHC during this data-taking period as well as the support staff from our institutions without whom ATLAS could not be operated efficiently.

We acknowledge the support of ANPCyT, Argentina; YerPhI, Armenia; ARC, Australia; BMWF, Austria; ANAS, Azerbaijan; SSTC, Belarus; CNPq and FAPESP, Brazil; NSERC, NRC and CFI, Canada; CERN; CONICYT, Chile; CAS, MOST and NSFC, China; COLCIENCIAS, Colombia; MSMT CR, MPO CR and VSC CR, Czech Republic; DNRF, DNSRC and Lundbeck Foundation, Denmark; ARTEMIS, European Union; IN2P3-CNRS, CEA-DSM/IRFU, France; GNAS, Georgia; BMBF, DFG, HGF, MPG and AvH Foundation, Germany; GSRT, Greece; ISF, MINERVA, GIF, DIP and Benoziyo Center, Israel; INFN, Italy; MEXT and JSPS, Japan; CNRST, Morocco; FOM and NWO, Netherlands; RCN, Norway;  MNiSW, Poland; GRICES and FCT, Portugal;  MERYS (MECTS), Romania;  MES of Russia and ROSATOM, Russian Federation; JINR; MSTD, Serbia; MSSR, Slovakia; ARRS and MVZT, Slovenia; DST/NRF, South Africa; MICINN, Spain; SRC and Wallenberg Foundation, Sweden; SER,  SNSF and Cantons of Bern and Geneva, Switzerland;  NSC, Taiwan; TAEK, Turkey; STFC, the Royal Society and Leverhulme Trust, United Kingdom; DOE and NSF, United States of America.

The crucial computing support from all WLCG partners is acknowledged gratefully, in particular from CERN and the ATLAS Tier-1 facilities at TRIUMF (Canada), NDGF (Denmark, Norway, Sweden), CC-IN2P3 (France), KIT/GridKA (Germany), INFN-CNAF (Italy), NL-T1 (Netherlands), PIC (Spain), ASGC (Taiwan), RAL (UK) and BNL (USA) and in the Tier-2 facilities worldwide.

\bibliographystyle{atlasnote}
\bibliography{MultiJets2011}

\newpage

\clearpage

\include{atlas_authlist}

\end{document}

%% file: susydefs.tex
\def\lsim{\mathrel{\rlap{\lower4pt\hbox{\hskip1pt$\sim$}}
    \raise1pt\hbox{$<$}}}                
\def\gsim{\mathrel{\rlap{\lower4pt\hbox{\hskip1pt$\sim$}}
    \raise1pt\hbox{$>$}}}                

\makeatletter
\renewcommand{\p@subfigure}{\arabic{figure}}
\renewcommand{\thesubfigure}{\alph{subfigure}}
\renewcommand{\@thesubfigure}{(\thesubfigure)}
\makeatother

%% file: atlas_authlist.tex
\begin{flushleft}
{\Large The ATLAS Collaboration}

\bigskip

G.~Aad$^{\rm 48}$,
B.~Abbott$^{\rm 111}$,
J.~Abdallah$^{\rm 11}$,
A.A.~Abdelalim$^{\rm 49}$,
A.~Abdesselam$^{\rm 118}$,
O.~Abdinov$^{\rm 10}$,
B.~Abi$^{\rm 112}$,
M.~Abolins$^{\rm 88}$,
H.~Abramowicz$^{\rm 153}$,
H.~Abreu$^{\rm 115}$
% ...
\bigskip

% ...
$^{10}$ Institute of Physics, Azerbaijan Academy of Sciences, Baku, Azerbaijan\\
$^{11}$ Institut de F\'isica d'Altes Energies and Departament de F\'isica de la Universitat Aut\`onoma  de Barcelona and ICREA, Barcelona, Spain\\
% ...
$^{48}$ Fakult\"{a}t f\"{u}r Mathematik und Physik, Albert-Ludwigs-Universit\"{a}t, Freiburg i.Br., Germany\\
$^{49}$ Section de Physique, Universit\'e de Gen\`eve, Geneva, Switzerland\\
% ...
$^{88}$ Department of Physics and Astronomy, Michigan State University, East Lansing MI, United States of America\\
% ...
$^{111}$ Homer L. Dodge Department of Physics and Astronomy, University of Oklahoma, Norman OK, United States of America\\
$^{112}$ Department of Physics, Oklahoma State University, Stillwater OK, United States of America\\
% ...
$^{115}$ LAL, Univ. Paris-Sud and CNRS/IN2P3, Orsay, France\\
% ...
$^{118}$ Department of Physics, Oxford University, Oxford, United Kingdom\\
% ...
$^{153}$ Raymond and Beverly Sackler School of Physics and Astronomy, Tel Aviv University, Tel Aviv, Israel\\
% ...
\end{flushleft}

%% file: MultiJets2011.bbl
\providecommand{\href}[2]{#2}\begingroup\raggedright\begin{thebibliography}{10}

\bibitem{Evans:2008zzb}
L.~Evans, (ed.~) and P.~Bryant, (ed.~)
\href{http://dx.doi.org/10.1088/1748-0221/3/08/S08001}{JINST {\bf 3} (2008)
  S08001}.

\bibitem{daCosta:2011qk}
{ATLAS} Collaboration,
  \href{http://dx.doi.org/10.1016/j.physletb.2011.05.061}{Phys. Lett. {\bf
  B701} (2011)  186--203},
\href{http://arxiv.org/abs/1102.5290}{{\tt arXiv:1102.5290}}.

\bibitem{Aad:2011ks}
{ATLAS} Collaboration,
  \href{http://dx.doi.org/10.1016/j.physletb.2011.06.015}{Phys. Lett. {\bf
  B701} (2011)  398--416},
\href{http://arxiv.org/abs/1103.4344}{{\tt arXiv:1103.4344}}.

\bibitem{Aad:2011xw}
{ATLAS} Collaboration, Submitted to Phys. Lett. B (2011)  ,
\href{http://arxiv.org/abs/1106.5327}{{\tt arXiv:1106.5327}}.

\bibitem{ATLAS-ZERO-LEPTON-2011}
{ATLAS} Collaboration, Submitted to Phys. Lett. B (2011)  ,
  \href{http://arxiv.org/abs/1109.6572}{{\tt arXiv:1109.6572}}.

\bibitem{Collaboration:2011ida}
{CMS} Collaboration,
\href{http://arxiv.org/abs/1106.4503}{{\tt arXiv:1106.4503}}.

\bibitem{Chatrchyan:2011ek}
{CMS} Collaboration,
\href{http://arxiv.org/abs/1107.1279}{{\tt arXiv:1107.1279}}.

\bibitem{Chatrchyan:2011nd}
{CMS} Collaboration,
\href{http://arxiv.org/abs/1106.4775}{{\tt arXiv:1106.4775}}.

\bibitem{susy_all_refs_no_titles}
{Yu.A. Golfand and E.P. Likhtman,
  \href{http://www.jetpletters.ac.ru/ps/1584/article_24309.shtml}{{JETP Lett.
  {\bf 13} (1971) 323-326}}.\newline A. Neveu and J.H. Schwartz,
  \href{http://dx.doi.org/10.1016/0550-3213(71)90448-2}{{Nucl. Phys. {\bf B31}
  (1971) 86-112}}.\newline A. Neveu and J.H. Schwartz,
  \href{http://dx.doi.org/10.1103/PhysRevD.4.1109}{{Phys. Rev. {\bf D4} (1971)
  1109-1111}}.\newline P. Ramond,
  \href{http://dx.doi.org/10.1103/PhysRevD.3.2415}{{Phys. Rev. {\bf D3} (1971)
  2415-2418}}.\newline D.V. Volkov and V.P. Akulov,
  \href{http://dx.doi.org/10.1016/0550-3213(72)90205-2}{{Phys. Lett. {\bf B46}
  (1973) 109-130}}.\newline J. Wess and B. Zumino,
  \href{http://dx.doi.org/10.1016/0550-3213(72)90584-6}{{Phys. Lett. {\bf B49}
  (1974) 52-60}}.\newline J. Wess and B. Zumino,
  \href{http://dx.doi.org/10.1016/0550-3213(74)90355-1}{{Nucl. Phys. {\bf B70}
  (1974) 39-50}}.}

\bibitem{Mangano:2002ea}
M.~L. Mangano, M.~Moretti, F.~Piccinini, R.~Pittau, and A.~D. Polosa
  \href{http://dx.doi.org/10.1088/1126-6708/2003/07/001}{JHEP {\bf 07} (2003)
  001},
\href{http://arxiv.org/abs/hep-ph/0206293}{{\tt arXiv:hep-ph/0206293}}.

\bibitem{Gleisberg:2008ta}
T.~Gleisberg et al. \href{http://dx.doi.org/10.1088/1126-6708/2009/02/007}{JHEP
  {\bf 02} (2009)  007},
\href{http://arxiv.org/abs/0811.4622}{{\tt arXiv:0811.4622}}.

\bibitem{Schumann:2007mg}
S.~Schumann and F.~Krauss
  \href{http://dx.doi.org/10.1088/1126-6708/2008/03/038}{JHEP {\bf 03} (2008)
  038},
\href{http://arxiv.org/abs/0709.1027}{{\tt arXiv:0709.1027}}.

\bibitem{Hoeche:2009rj}
S.~Hoeche, F.~Krauss, S.~Schumann, and F.~Siegert
  \href{http://dx.doi.org/10.1088/1126-6708/2009/05/053}{JHEP {\bf 05} (2009)
  053},
\href{http://arxiv.org/abs/0903.1219}{{\tt arXiv:0903.1219}}.

\bibitem{Collaboration:2011hh}
{ATLAS} Collaboration,
  \href{http://dx.doi.org/10.1103/PhysRevLett.106.131802}{Phys. Rev. Lett. {\bf
  102} (2011)  131802},
\href{http://arxiv.org/abs/1102.2357}{{\tt arXiv:1102.2357}}.

\bibitem{Aad:2011xk}
{ATLAS} Collaboration,
  \href{http://dx.doi.org/10.1140/epjc/s10052-011-1647-9}{Eur. Phys. J. {\bf
  C71} (2011)  1647},
\href{http://arxiv.org/abs/1103.6208}{{\tt arXiv:1103.6208}}.

\bibitem{ATLAS-ONE-LEPTON-SUSY11}
{ATLAS} Collaboration, Submitted to Physical Review D (2011)  ,
  \href{http://arxiv.org/abs/1109.6606}{{\tt arXiv:1109.6606}}.

\bibitem{Aad:2011kz}
{ATLAS} Collaboration, Submitted to EPJC (2011)  ,
\href{http://arxiv.org/abs/1107.0561}{{\tt arXiv:1107.0561}}.

\bibitem{Chamseddine:1982jx}
A.~H. Chamseddine, R.~L. Arnowitt, and P.~Nath
  \href{http://dx.doi.org/10.1103/PhysRevLett.49.970}{Phys.Rev.Lett. {\bf 49}
  (1982)  970}.

\bibitem{Barbieri:1982eh}
R.~Barbieri, S.~Ferrara, and C.~A. Savoy
\href{http://dx.doi.org/10.1016/0370-2693(82)90685-2}{Phys. Lett. {\bf B119}
  (1982)  343}.

\bibitem{Ibanez:1982ee}
L.~E. {Ibanez} \href{http://dx.doi.org/10.1016/0370-2693(82)90604-9}{Phys.Lett.
  {\bf B118} (1982)  73}.

\bibitem{Hall:1983iz}
L.~J. Hall, J.~D. Lykken, and S.~Weinberg
  \href{http://dx.doi.org/10.1103/PhysRevD.27.2359}{Phys.Rev. {\bf D27} (1983)
  2359--2378}.

\bibitem{Ohta:1982wn}
N.~{Ohta} \href{http://dx.doi.org/10.1143/PTP.70.542}{Prog.Theor.Phys. {\bf 70}
  (1983)  542}.

\bibitem{Kane:1993td}
G.~L. Kane, C.~F. Kolda, L.~Roszkowski, and J.~D. Wells
  \href{http://dx.doi.org/10.1103/PhysRevD.49.6173}{Phys.Rev. {\bf D49} (1994)
  6173--6210}, \href{http://arxiv.org/abs/hep-ph/9312272}{{\tt
  arXiv:hep-ph/9312272}}.

\bibitem{Aad:2008zzm}
{ATLAS} Collaboration,
\href{http://dx.doi.org/10.1088/1748-0221/3/08/S08003}{JINST {\bf 3} (2008)
  S08003}.

\bibitem{Cacciari:2008gp}
M.~Cacciari, G.~P. Salam, and G.~Soyez
  \href{http://dx.doi.org/10.1088/1126-6708/2008/04/063}{JHEP {\bf 04} (2008)
  063},
\href{http://arxiv.org/abs/0802.1189}{{\tt arXiv:0802.1189}}.

\bibitem{Cacciari:2005hq}
M.~Cacciari and G.~P. Salam
  \href{http://dx.doi.org/10.1016/j.physletb.2006.08.037}{Phys. Lett. {\bf
  B641} (2006)  57--61},
\href{http://arxiv.org/abs/hep-ph/0512210}{{\tt arXiv:hep-ph/0512210}}.

\bibitem{topoclusters}
W.~Lampl et al.
  \href{http://cdsweb.cern.ch/record/1099735}{{ATL-LARG-PUB-2008-002}}, 2008.

\bibitem{atlas-jetcal}
{ATLAS} Collaboration,
  \href{http://cdsweb.cern.ch/record/1281329}{{ATLAS-CONF-2010-056}}, Jul,
  2010.

\bibitem{ATLAS-CONF-2011-102}
{ATLAS} Collaboration,
  \href{http://cdsweb.cern.ch/record/1369219}{{ATLAS-CONF-2011-102}}, Jul,
  2011.

\bibitem{Aad:2010yt}
{ATLAS} Collaboration, \href{http://dx.doi.org/10.1007/JHEP12(2010)060}{JHEP
  {\bf 12} (2010)  060},
\href{http://arxiv.org/abs/1010.2130}{{\tt arXiv:1010.2130}}.

\bibitem{atlas-jet-clean}
{ATLAS} Collaboration,
  \href{http://cdsweb.cern.ch/record/1277678}{{ATLAS-CONF-2010-038}}, Jul,
  2010.

\bibitem{Collaboration:2011tq}
{ATLAS} Collaboration, Submitted to European Physics Journal C (2011)  ,
\href{http://arxiv.org/abs/1107.2092}{{\tt arXiv:1107.2092}}.

\bibitem{Agostinelli:2002hh}
{GEANT4} Collaboration, S.~Agostinelli et al.
\href{http://dx.doi.org/10.1016/S0168-9002(03)01368-8}{Nucl. Instrum. Meth.
  {\bf A506} (2003)  250--303}.

\bibitem{:2010wqa}
{ATLAS Collaboration} Collaboration,
  \href{http://dx.doi.org/10.1140/epjc/s10052-010-1429-9}{Eur.Phys.J. {\bf C70}
  (2010)  823--874}, \href{http://arxiv.org/abs/1005.4568}{{\tt
  arXiv:1005.4568}}.

\bibitem{Pumplin:2002vw}
J.~Pumplin et al. \href{http://dx.doi.org/10.1088/1126-6708/2002/07/012}{JHEP
  {\bf 07} (2002)  012},
\href{http://arxiv.org/abs/hep-ph/0201195}{{\tt arXiv:hep-ph/0201195}}.

\bibitem{Corcella:2000bw}
G.~Corcella et al. \href{http://dx.doi.org/10.1088/1126-6708/2001/01/010}{JHEP
  {\bf 01} (2001)  010},
\href{http://arxiv.org/abs/hep-ph/0011363}{{\tt arXiv:hep-ph/0011363}}.

\bibitem{herwig65long}
G.~Corcella et al., 2002.
\newblock \href{http://arxiv.org/abs/hep-ph/0210213}{{\tt
  arXiv:hep-ph/0210213}}.

\bibitem{Butterworth:1996zw}
J.~M. Butterworth, J.~R. Forshaw, and M.~H. Seymour
  \href{http://dx.doi.org/10.1007/s002880050286}{Z. Phys. {\bf C72} (1996)
  637--646},
\href{http://arxiv.org/abs/hep-ph/9601371}{{\tt arXiv:hep-ph/9601371}}.

\bibitem{Ellis:1985vn}
S.~D. Ellis, R.~Kleiss, and W.~J. Stirling
\href{http://dx.doi.org/10.1016/0370-2693(85)90425-3}{Phys. Lett. {\bf B154}
  (1985)  435}.

\bibitem{Berends:1989cf}
F.~A. Berends, W.~T. Giele, H.~Kuijf, R.~Kleiss, and W.~J. Stirling
\href{http://dx.doi.org/10.1016/0370-2693(89)91081-2}{Phys. Lett. {\bf B224}
  (1989)  237}.

\bibitem{Berger:2010gf}
C.~F. Berger et al.
  \href{http://dx.doi.org/10.1016/j.nuclphysbps.2010.08.025}{Nucl. Phys. Proc.
  Suppl. {\bf 205-206} (2010)  92--97},
\href{http://arxiv.org/abs/1005.3728}{{\tt arXiv:1005.3728}}.

\bibitem{Abazov:2002gy}
{D0} Collaboration, V.~M. Abazov et al.
  \href{http://dx.doi.org/10.1103/PhysRevD.67.012004}{Phys. Rev. {\bf D67}
  (2003)  012004},
\href{http://arxiv.org/abs/hep-ex/0205019}{{\tt arXiv:hep-ex/0205019}}.

\bibitem{Wagner:2003cv}
{CDF} Collaboration, W.~Wagner
  \href{http://dx.doi.org/10.1140/epjcd/s2003-03-120-7}{Eur. Phys. J. {\bf C33}
  (2004)  s238--s240},
\href{http://arxiv.org/abs/hep-ex/0312008}{{\tt arXiv:hep-ex/0312008}}.

\bibitem{:2010pg}
{ATLAS} Collaboration,
  \href{http://dx.doi.org/10.1016/j.physletb.2011.03.012}{Phys. Lett. {\bf
  B698} (2011)  325--345},
\href{http://arxiv.org/abs/1012.5382}{{\tt arXiv:1012.5382}}.

\bibitem{Gleisberg:2008fv}
T.~Gleisberg and S.~Hoeche
  \href{http://dx.doi.org/10.1088/1126-6708/2008/12/039}{JHEP {\bf 12} (2008)
  039},
\href{http://arxiv.org/abs/0808.3674}{{\tt arXiv:0808.3674}}.

\bibitem{isajet}
{F.E.~Paige et al.,
  \href{http://xxx.lanl.gov/abs/hep-ph/0312045v1}{arXiv:hep-ph/0312045}
  (2003)}.

\bibitem{Bahr:2008pv}
M.~Bahr et al. \href{http://dx.doi.org/10.1140/epjc/s10052-008-0798-9}{Eur.
  Phys. J. {\bf C58} (2008)  639--707},
\href{http://arxiv.org/abs/0803.0883}{{\tt arXiv:0803.0883}}.

\bibitem{Beenakker:1996ch}
W.~Beenakker, R.~Hopker, M.~Spira, and P.~M. Zerwas
  \href{http://dx.doi.org/10.1016/S0550-3213(97)00084-9}{Nucl. Phys. {\bf B492}
  (1997)  51--103},
\href{http://arxiv.org/abs/hep-ph/9610490}{{\tt arXiv:hep-ph/9610490}}.

\bibitem{Nadolsky:2008zw}
P.~M. Nadolsky et al. \href{http://dx.doi.org/10.1103/PhysRevD.78.013004}{Phys.
  Rev. {\bf D78} (2008)  013004},
\href{http://arxiv.org/abs/0802.0007}{{\tt arXiv:0802.0007}}.

\bibitem{lumi_summer_11}
{ATLAS} Collaboration,
  \href{http://cdsweb.cern.ch/record/1376384}{{ATLAS-CONF-2011-116}}, Jul,
  2011.

\bibitem{Aad:2011dr}
{ATLAS} Collaboration,
  \href{http://dx.doi.org/10.1140/epjc/s10052-011-1630-5}{Eur. Phys. J. {\bf
  C71} (2011)  1630},
\href{http://arxiv.org/abs/1101.2185}{{\tt arXiv:1101.2185}}.

\bibitem{clsread}
A.Read \href{http://dx.doi.org/10.1088/0954-3899/28/10/313}{Journal of Physics
  G: Nucl. Part. Phys. {\bf 28} (2002)  2693--2704}.

\bibitem{multijet_2011_auxiliary}
{{\url{{http://hepdata.cedar.ac.uk/view/red4931}}}}.

\bibitem{Abachi:1995ng}
{D0} Collaboration,
\href{http://dx.doi.org/10.1103/PhysRevLett.75.618}{Phys. Rev. Lett. {\bf 75}
  (1995)  618--623}.

\bibitem{Affolder:2001tc}
{CDF} Collaboration,
  \href{http://dx.doi.org/10.1103/PhysRevLett.88.041801}{Phys. Rev. Lett. {\bf
  88} (2002)  041801},
\href{http://arxiv.org/abs/hep-ex/0106001}{{\tt arXiv:hep-ex/0106001}}.

\bibitem{Aaltonen:2008rv}
{{CDF}} Collaboration,
  \href{http://dx.doi.org/10.1103/PhysRevLett.102.121801}{Phys. Rev. Lett. {\bf
  102} (2009)  121801},
\href{http://arxiv.org/abs/0811.2512}{{\tt arXiv:0811.2512}}.

\bibitem{:2007ww}
{{D0}} Collaboration,
  \href{http://dx.doi.org/10.1016/j.physletb.2008.01.042}{Phys. Lett. {\bf
  B660} (2008)  449--457},
\href{http://arxiv.org/abs/0712.3805}{{\tt arXiv:0712.3805}}.

\bibitem{lepsusy}
{LEP SUSY Working Group (ALEPH, DELPHI, L3, OPAL), Notes LEPSUSYWG/01-03.1 and
  04-01.1, \url{http://lepsusy.web.cern.ch/lepsusy/Welcome.html}}.

\end{thebibliography}\endgroup
